\documentclass[a4paper,11pt]{article}
\usepackage{jheppub} 
\usepackage{hyperref}
\usepackage[T1]{fontenc} 

\title{\boldmath Muon conversion to electron in nuclei within the $\mu\nu$SSM with a 125 GeV Higgs}

\author[a,b]{Hai-Bin Zhang,}
\author[a,b]{Tai-Fu Feng,}
\author[b]{Guo-Hui Luo,}
\author[a]{Zhao-Feng Ge}
\author[a]{and Shu-Min Zhao}

\affiliation[a]{Department of Physics, Hebei University, \\ Baoding, 071002,  China}
\affiliation[b]{Department of Physics, Dalian University of Technology, \\
Dalian, 116024, China}

\emailAdd{hbzhang@mail.dlut.edu.cn}
\emailAdd{fengtf@hbu.edu.cn}
\emailAdd{ghuiluo@gmail.com}
\emailAdd{algezhaofeng@126.com}
\emailAdd{smzhao@hbu.edu.cn}

\abstract{Within framework of the $\mu$ from $\nu$ Supersymmetric Standard Model ($\mu\nu$SSM), three exotic right-handed neutrino superfields induce new sources for lepton-flavor violation.  In this work, we investigate muon conversion to electron in nuclei within the $\mu\nu$SSM in detail. With a 125 GeV Higgs, the numerical results indicate that the $\mu-e$ conversion rates in nuclei within the $\mu\nu$SSM can reach the experimental upper bound, which could be detected with the future experimental sensitivities.}


\arxivnumber{1305.4352}

\begin{document}
\maketitle
\flushbottom

\section{Introduction}
\label{intro}
Lepton-flavor violation (LFV) is a window of new physics beyond the Standard Model (SM), because the lepton-flavor number is conserved in the Standard Model. Among the various candidates for new physics that produce potentially observable
effects in LFV processes, one of the most appealing are Supersymmetric (SUSY)
extensions of the SM, which can violate lepton number naturally. In SUSY extensions of the SM, the R-parity of a particle is defined as $R = (-1)^{L+3B+2S}$~\cite{Rp1,Rp2} and can be violated if either the lepton number~($L$) or baryon number~($B$) is not conserved~\cite{RPV,R1,R2,R3,R4,R5,R6,R7,R8,R9,R10,R11,R12,R13}, where $S$ denotes the spin of concerned component field. Note that $R=+1$ for particles and $-1$ for superparticles.

Differing from the models in refs.~\cite{RPV,R1,R2,R3,R4,R5,R6,R7,R8,R9,R10,R11,R12,R13}, the authors of refs.~\cite{mnSSM,mnSSM1,mnSSM2} propose a SUSY extension of the SM named as the ``$\mu$ from $\nu$ Supersymmetric Standard Model''~($\mu\nu$SSM), which solves the $\mu$ problem~\cite{m-problem} of the Minimal Supersymmetric Standard Model (MSSM)~\cite{MSSM1,MSSM2,MSSM3,MSSM4} through the lepton number  breaking couplings between the right-handed neutrino superfields and the Higgses  $\epsilon _{ab}{\lambda _i}\hat \nu _i^c\hat H_d^a\hat H_u^b$ in the superpotential. The effective $\mu$ term $\epsilon _{ab} \mu \hat H_d^a\hat H_u^b$ is generated spontaneously through right-handed sneutrino vacuum expectation values (VEVs), $\mu  = {\lambda _i}\left\langle {\tilde \nu _i^c} \right\rangle$, as the electroweak symmetry is broken (EWSB). Largely differing from the other models~\cite{RPV,R1,R2,R3,R4,R5,R6,R7,R8,R9,R10,R11,R12,R13,MSSM1,MSSM2,MSSM3,MSSM4}, the $\mu\nu$SSM introduces three exotic right-handed sneutrinos $\hat{\nu}_i^c$, and once EWSB the right-handed sneutrinos give nonzero VEVs. In addition, the nonzero VEVs of right-handed sneutrinos induce new sources for lepton-flavor violation. In our previous work, we had analyzed some LFV processes $l_j^-\rightarrow l_i^-\gamma$ and $l_j^-  \rightarrow l_i^- l_i^- l_i^+$ in the $\mu\nu$SSM~\cite{Zhang}. In this work, we continue to analyzed the LFV processes on muon conversion to electron in nuclei within the $\mu\nu$SSM. The $\mu-e$ conversion rates have been calculated in the literature for the various possible types of seesaw, with right-handed neutrinos~\cite{ue1,ue2,ue3,ue4,ue5,uec,ue6,ue7,ueI,ue8}, scalar triplet(s)~\cite{ue9,ue10,ue11} and fermion triplets~\cite{ue12}. In the $\mu\nu$SSM, three neutrino masses can be generated at the tree level through the mixing with the neutralinos including three right-handed neutrinos~\cite{neutrino-mass,neu-mass1,neu-mass2,neu-mass3}.

Recently neutrino experiments develop quickly, which give the constraints on the parameters. If the left-handed scalar neutrinos acquire nonzero VEVs when the electroweak symmetry is broken, the tiny neutrino masses are aroused~\cite{Feng2} to account for the experimental data on neutrino oscillations~\cite{oscillation1,oscillation2,oscillation3}.
Three flavor neutrinos $\nu_{e,\mu,\tau}$ are mixed into three massive neutrinos $\nu_{1,2,3}$ during their flight, and the mixings are described by the Pontecorvo-Maki-Nakagawa-Sakata unitary matrix
$U_{_{PMNS}}$~\cite{neutrino-oscillations,neutrino-oscillations1}. Through the
several recent reactor oscillation experiments~\cite{theta13,theta13-1,theta13-2,theta13-3,theta13-4}, $\theta_{13}$ is now precisely known. The global fit of $\theta_{13}$ gives~\cite{Garcia}
\begin{eqnarray}
\sin^2\theta_{13}=0.023\pm 0.0023.
\label{neutrino-oscillations1}
\end{eqnarray}
And the other experimental observations of the parameters in $U_{_{PMNS}}$ for the normal mass hierarchy~\cite{Valle1} show that~\cite{PDG}
\begin{eqnarray}
&&\;\;\Delta m_{21}^2 =7.58_{-0.26}^{+0.22}\times 10^{-5} {\rm eV}^2,\nonumber\\
&&\;\;\Delta m_{32}^2 =2.35_{-0.09}^{+0.12}\times 10^{-3} {\rm eV}^2,\nonumber\\
&&\sin^2\theta_{12} =0.306_{-0.015}^{+0.018},\qquad \sin^2\theta_{23}=0.42_{-0.03}^{+0.08}.
\label{neutrino-oscillations2}
\end{eqnarray}

Here, we can use the neutrino oscillation experimental data to restrain the input parameters in the $\mu\nu{\rm SSM}$. Lately, a neutral Higgs with mass $m_{h}\sim 124-126\;{\rm GeV}$ reported by ATLAS~\cite{ATLAS} and CMS~\cite{CMS} also contributes a strict constraint on relevant parameter space of the model.  Then, we analyze the $\mu-e$ conversion rates within the $\mu\nu$SSM with the 125 GeV Higgs. As compared to the present sensitivities of the $\mu-e$ conversion rates in different nuclei~\cite{CRTi,CRAu,CRPb}
\begin{eqnarray}
&&\;\;\,{\rm{CR}}(\mu \to e:{_{22}^{48}{\rm{Ti}}}) < 4.3\times 10^{-12},\nonumber\\
&&{\rm{CR}}(\mu \to e:{_{\:79}^{197}{\rm{Au}}}) < 7\times 10^{-13},\nonumber\\
&&\,{\rm{CR}}(\mu \to e:{_{\:82}^{207}{\rm{Pb}}}) < 4.6\times 10^{-11},
\label{CRpresent}
\end{eqnarray}
the numerical results indicate that the new physics contributes large corrections to the $\mu-e$ conversion rates in some parameter space of the model. Recently, the MEG experiment updates a new upper limit on the branching ratio of LFV process $\mu\rightarrow e\gamma$~\cite{MEG}
\begin{eqnarray}
{\rm{Br}}(\mu\rightarrow e\gamma)<5.7\times10^{-13},
\end{eqnarray}
which is a four times more stringent than the previous limit~\cite{MEG1}. The new upper limit also gives strong constraint on the $\mu\nu$SSM.

The outline of the paper is as follow. In section~\ref{sec:2}, we present the ingredients of the $\mu\nu$SSM by introducing its superpotential and the general soft SUSY-breaking terms, in particular the unphysical Goldstone bosons are strictly separated from the scalars. Section~\ref{sec:3} gives the radiative correction to the SM-like Higgs. In section~\ref{sec:4}, we analyze the muon conversion to electron in nuclei within the $\mu\nu$SSM. The numerical analysis is given in section~\ref{sec:5}, and the conclusions are summarized in section~\ref{sec:6}. The tedious formulae are collected in appendices~\ref{appendix-mini}--\ref{appendix-integral}.

\section{The $\mu\nu$SSM\label{sec:2}}
Besides the superfields of the MSSM, the $\mu\nu$SSM introduces three singlet right-handed neutrino superfields $\hat{\nu}_i^c$. The corresponding superpotential of the $\mu\nu$SSM is given by~\cite{mnSSM}
\begin{eqnarray}
&&W \:=\:{\epsilon _{ab}}\left( {{Y_{{u_{ij}}}}\hat H_u^b\hat Q_i^a\hat u_j^c + {Y_{{d_{ij}}}}\hat H_d^a\hat Q_i^b\hat d_j^c
+ {Y_{{e_{ij}}}}\hat H_d^a\hat L_i^b\hat e_j^c + {Y_{{\nu _{ij}}}}\hat H_u^b\hat L_i^a\hat \nu _j^c} \right)  \nonumber\\
&&\qquad\quad - \: {\epsilon _{ab}}{\lambda _i}\hat \nu _i^c\hat H_d^a\hat H_u^b + \frac{1}{3}{\kappa _{ijk}}\hat \nu _i^c\hat \nu _j^c\hat \nu _k^c ,
\end{eqnarray}
where $\hat H_u^T = \Big( {\hat H_u^ + ,\hat H_u^0} \Big)$, $\hat H_d^T = \Big( {\hat H_d^0,\hat H_d^ - } \Big)$, $\hat Q_i^T = \Big( {{{\hat u}_i},{{\hat d}_i}} \Big)$, $\hat L_i^T = \Big( {{{\hat \nu}_i},{{\hat e}_i}} \Big)$ are $SU(2)$ doublet superfields, and $\hat u_j^c$, $\hat d_j^c$ and $\hat e_j^c$ represent the singlet up-type quark, down-type quark and lepton superfields, respectively.  In addition, $Y_{u,d,e,\nu}$, $\lambda$ and $\kappa$ are dimensionless matrices, a vector and a totally symmetric tensor.  $a,b$ are SU(2) indices with antisymmetric tensor $\epsilon_{12}=-\epsilon_{21}=1$, and $i,j,k=1,\;2,\;3$. The summation convention is implied on repeated indices in the following.

In the superpotential, the first three terms are almost the same as the MSSM. Next two terms can generate the effective bilinear terms $\epsilon _{ab} \varepsilon_i \hat H_u^b\hat L_i^a$, $\epsilon _{ab} \mu \hat H_d^a\hat H_u^b$,  and $\varepsilon_i= Y_{\nu _{ij}} \left\langle {\tilde \nu _j^c} \right\rangle$, $\mu  = {\lambda _i}\left\langle {\tilde \nu _i^c} \right\rangle$,  once the electroweak symmetry is broken. The last term generates the effective Majorana masses for neutrinos at the electroweak scale. And the last two terms explicitly violate lepton number and R-parity.

The general soft SUSY-breaking terms in the $\mu\nu$SSM are given as
\begin{eqnarray}
&&- \mathcal{L}_{soft}\:=\:m_{{{\tilde Q}_{ij}}}^{\rm{2}}\tilde Q{_i^{a\ast}}\tilde Q_j^a
+ m_{\tilde u_{ij}^c}^{\rm{2}}\tilde u{_i^{c\ast}}\tilde u_j^c + m_{\tilde d_{ij}^c}^2\tilde d{_i^{c\ast}}\tilde d_j^c
+ m_{{{\tilde L}_{ij}}}^2\tilde L_i^{a\ast}\tilde L_j^a  \nonumber\\
&&\hspace{1.8cm} + \: m_{\tilde e_{ij}^c}^2\tilde e{_i^{c\ast}}\tilde e_j^c + m_{{H_d}}^{\rm{2}} H_d^{a\ast} H_d^a
+ m_{{H_u}}^2H{_u^{a\ast}}H_u^a + m_{\tilde \nu_{ij}^c}^2\tilde \nu{_i^{c\ast}}\tilde \nu_j^c \nonumber\\
&&\hspace{1.8cm}  + \: \epsilon_{ab}{\left[{{({A_u}{Y_u})}_{ij}}H_u^b\tilde Q_i^a\tilde u_j^c
+ {{({A_d}{Y_d})}_{ij}}H_d^a\tilde Q_i^b\tilde d_j^c + {{({A_e}{Y_e})}_{ij}}H_d^a\tilde L_i^b\tilde e_j^c + {\rm{H.c.}} \right]} \nonumber\\
&&\hspace{1.8cm}  + \left[ {\epsilon _{ab}}{{({A_\nu}{Y_\nu})}_{ij}}H_u^b\tilde L_i^a\tilde \nu_j^c
- {\epsilon _{ab}}{{({A_\lambda }\lambda )}_i}\tilde \nu_i^c H_d^a H_u^b
+ \frac{1}{3}{{({A_\kappa }\kappa )}_{ijk}}\tilde \nu_i^c\tilde \nu_j^c\tilde \nu_k^c + {\rm{H.c.}} \right] \nonumber\\
&&\hspace{1.8cm}  - \: \frac{1}{2}\left({M_3}{{\tilde \lambda }_3}{{\tilde \lambda }_3}
+ {M_2}{{\tilde \lambda }_2}{{\tilde \lambda }_2} + {M_1}{{\tilde \lambda }_1}{{\tilde \lambda }_1} + {\rm{H.c.}} \right).
\end{eqnarray}
Here, the front two lines consist of squared-mass terms of squarks, sleptons and Higgses. The next two lines contain the trilinear scalar couplings. In the last line, $M_3$, $M_2$ and $M_1$ denote Majorana masses corresponding to $SU(3)$, $SU(2)$ and $U(1)$ gauginos $\hat{\lambda}_3$, $\hat{\lambda}_2$ and $\hat{\lambda}_1$, respectively. In addition to the terms from $\mathcal{L}_{soft}$, the tree-level scalar potential receives the usual D and F term contributions~\cite{mnSSM1}.

When the electroweak symmetry is spontaneously broken, the neutral scalars develop in general the vacuum expectation values:
\begin{eqnarray}
\langle H_d^0 \rangle = \upsilon_d , \qquad \langle H_u^0 \rangle = \upsilon_u , \qquad
\langle \tilde \nu_i \rangle = \upsilon_{\nu_i} , \qquad \langle \tilde \nu_i^c \rangle = \upsilon_{\nu_i^c} .
\end{eqnarray}
Thus one can define neutral scalars as usual
\begin{eqnarray}
&&H_d^0=\frac{h_d + i P_d}{\sqrt{2}} + \upsilon_d, \qquad\; \tilde \nu_i = \frac{(\tilde \nu_i)^\Re + i (\tilde \nu_i)^\Im}{\sqrt{2}} + \upsilon_{\nu_i},  \nonumber\\
&&H_u^0=\frac{h_u + i P_u}{\sqrt{2}} + \upsilon_u, \qquad \tilde \nu_i^c = \frac{(\tilde \nu_i^c)^\Re + i (\tilde \nu_i^c)^\Im}{\sqrt{2}} + \upsilon_{\nu_i^c}.
\end{eqnarray}
And one can have
\begin{eqnarray}
\tan\beta={\upsilon_u\over\sqrt{\upsilon_d^2+\upsilon_{\nu_i}\upsilon_{\nu_i}}}.
\label{tanb}
\end{eqnarray}

In the following, we will assume that all parameters in the potential are real for simplicity. After EWSB, the scalar mass matrices $M_S^2$, $M_P^2$, $M_{S^{\pm}}^2$, $M_{\tilde{u}}^2$ and $M_{\tilde{d}}^2$ are given in appendix~\ref{appendix-mass}. Making use of the minimization conditions of the tree-level neutral scalar potential, which are given in appendix~\ref{appendix-mini}, the CP-odd neutral scalar mass matrix $M_P^2$ and charged scalar mass matrix $M_{S^{\pm}}^2$ can respectively isolate massless unphysical Goldstone bosons $G^0$ and $G^{\pm}$, which can be written as~\cite{Zhang}
\begin{eqnarray}
G^0 = {1 \over \sqrt{\upsilon_d^2+\upsilon_u^2+\upsilon_{\nu_i} \upsilon_{\nu_i}}} \Big(\upsilon_d {P_d}-\upsilon_u{P_u}+\upsilon_{\nu_i}{(\tilde \nu_i)^\Im}\Big)
\end{eqnarray}
and
\begin{eqnarray}
G^{\pm} = {1 \over \sqrt{\upsilon_d^2+\upsilon_u^2+\upsilon_{\nu_i} \upsilon_{\nu_i}}} \Big(\upsilon_d H_d^{\pm} - \upsilon_u {H_u^{\pm}}+\upsilon_{\nu_i}\tilde e_{L_i}^{\pm}\Big)
\end{eqnarray}
through an $8\times8$ matrix $Z_H$
\begin{eqnarray}
Z_H = \left( {\begin{array}{*{20}{c}}
   \frac{\upsilon_d}{\upsilon_{_{\rm{EW}}}} & \frac{\upsilon_u}{\upsilon_{_{\rm{SM}}}} & \frac{\upsilon_{\nu_1}\upsilon_d}{\upsilon_{_{\rm{EW}}}\upsilon_{_{\rm{SM}}}} & \frac{\upsilon_{\nu_2}\upsilon_d}{\upsilon_{_{\rm{EW}}}\upsilon_{_{\rm{SM}}}} & \frac{\upsilon_{\nu_3}\upsilon_d}{\upsilon_{_{\rm{EW}}}\upsilon_{_{\rm{SM}}}} & 0_{1\times3}   \\  [6pt]
   -\frac{\upsilon_u}{\upsilon_{_{\rm{EW}}}} & \frac{\upsilon_d}{\upsilon_{_{\rm{SM}}}} & -\frac{\upsilon_{\nu_1}\upsilon_u}{\upsilon_{_{\rm{EW}}}\upsilon_{_{\rm{SM}}}} & -\frac{\upsilon_{\nu_2}\upsilon_u}{\upsilon_{_{\rm{EW}}}\upsilon_{_{\rm{SM}}}} & -\frac{\upsilon_{\nu_3}\upsilon_u}{\upsilon_{_{\rm{EW}}}\upsilon_{_{\rm{SM}}}} & 0_{1\times3}   \\  [6pt]
   \frac{\upsilon_{\nu_1}}{\upsilon_{_{\rm{EW}}}} & 0 & -\frac{\upsilon_{_{\rm{SM}}}}{\upsilon_{_{\rm{EW}}}} & \frac{\upsilon_{\nu_3}}{\upsilon_{_{\rm{EW}}}} & -\frac{\upsilon_{\nu_2}}{\upsilon_{_{\rm{EW}}}} & 0_{1\times3} \\  [6pt]
   \frac{\upsilon_{\nu_2}}{\upsilon_{_{\rm{EW}}}} & 0 & -\frac{\upsilon_{\nu_3}}{\upsilon_{_{\rm{EW}}}} & -\frac{\upsilon_{_{\rm{SM}}}}{\upsilon_{_{\rm{EW}}}} & \frac{\upsilon_{\nu_1}}{\upsilon_{_{\rm{EW}}}} & 0_{1\times3}  \\  [6pt]
   \frac{\upsilon_{\nu_3}}{\upsilon_{_{\rm{EW}}}} & 0 & \frac{\upsilon_{\nu_2}}{\upsilon_{_{\rm{EW}}}} & -\frac{\upsilon_{\nu_1}}{\upsilon_{_{\rm{EW}}}} & -\frac{\upsilon_{_{\rm{SM}}}}{\upsilon_{_{\rm{EW}}}} & 0_{1\times3}  \\  [6pt]
   0_{3\times1} & 0_{3\times1} & 0_{3\times1} & 0_{3\times1} & 0_{3\times1} & 1_{3\times3}  \\  [6pt]
\end{array}} \right),
\end{eqnarray}
where $\upsilon_{_{\rm{SM}}}=\sqrt{\upsilon_d^2+\upsilon_u^2}$ and $\upsilon_{_{\rm{EW}}}=\sqrt{\upsilon_d^2+\upsilon_u^2+\upsilon_{\nu_i}\upsilon_{\nu_i}}$. Here we can check that the matrix $Z_H$ is unitary, $Z_H^T Z_H=Z_H Z_H^T=1$. In the physical (unitary) gauge, the Goldstone bosons $G^0$ and $G^{\pm}$ are eaten by $Z$-boson and $W$-boson, respectively, and disappear from the Lagrangian.
And the masses of neutral and charged gauge bosons are given by
\begin{eqnarray}
\left\{ {\begin{array}{l}
    \:m_{_Z}={e\over {\sqrt{2}s_{_W} c_{_W}}}\sqrt{\upsilon_u^2+\upsilon_d^2+\upsilon_{\nu_i} \upsilon_{\nu_i}},  \\ [6pt]
    m_{_W}={e\over\sqrt{2}s_{_W}}\sqrt{\upsilon_u^2+\upsilon_d^2+\upsilon_{\nu_i} \upsilon_{\nu_i}}, \\ [6pt]
\end{array}} \right.
\end{eqnarray}
where $e$ is the electromagnetic coupling constant, $s_{_W}=\sin\theta_{_W}$ and $c_{_W}=\cos\theta_{_W}$ with $\theta_{_W}$ denoting the Weinberg angle, respectively.

\section{Radiative correction to the SM-like Higgs\label{sec:3}}

In the $\mu\nu$SSM, left and right-handed sneutrino VEVs lead to mixing of the neutral components of the Higgs doublets with the sneutrinos producing an $8\times8$ CP-even  neutral scalar mass matrix, which can be found in appendix~\ref{appendix-mass}. Neglecting the terms containing small coupling $Y_{\nu_i} \sim \mathcal{O}(10^{-7})$ and $\upsilon_{\nu_i} \sim \mathcal{O}(10^{-4}{\rm{GeV}})$, and implying the condition  \cite{mnSSM2}
\begin{eqnarray}
A_{\lambda_i}=\frac{{2\mu }}{\sin 2\beta}-\frac{2}{\lambda_i} \sum\limits_{j,k} \kappa_{ijk}\lambda_j \upsilon_{\nu_k^c},
\end{eqnarray}
where $\mu=\lambda_i \upsilon_{\nu_i^c}$, the $8\times8$ CP-even neutral scalar mass matrix can be decoupled from the $2\times2$  Higgs doublet submatrix. However, the condition is sufficient but not necessary. If the off-diagonal mixing terms of the CP-even neutral scalar mass matrix are enough smaller than the diagonal terms, the contribution of the off-diagonal mixing terms to the diagonal SM-like Higgs mass is small, which can be neglected. Actually, we will use this mechanism in the numerical calculation.

It is well known since quite some time that radiative corrections modify the tree level mass squared matrix of neutral Higgs substantially in the MSSM, with the main effect from loops involving the top quark and its scalar partner $\tilde{t}_{1,2}$~\cite{Haber}. In order to obtain masses of the neutral doublet-like Higgs reasonably, we consider the dominating radiative corrections from the third fermions and corresponding supersymmetric partners in the $\mu\nu$SSM. The $2\times2$ $\tilde{t}_L-\tilde{t}_R$, $\tilde{b}_L-\tilde{b}_R$ and $\tilde{\tau}_L-\tilde{\tau}_R$ mass squared matrices respectively are
\begin{eqnarray}
M_{\tilde{t}}^2=\left(\begin{array}{ll}M_{\tilde{u}_{L_3 L_3}}^2&M_{\tilde{u}_{L_3 R_3}}^2\\
M_{\tilde{u}_{L_3 R_3}}^2&M_{\tilde{u}_{R_3 R_3}}^2\end{array}\right),\qquad
M_{\tilde{b}}^2=\left(\begin{array}{ll}M_{\tilde{d}_{L_3 L_3}}^2&M_{\tilde{d}_{L_3 R_3}}^2\\
M_{\tilde{d}_{L_3 R_3}}^2&M_{\tilde{d}_{R_3 R_3}}^2\end{array}\right)
\end{eqnarray}
and
\begin{eqnarray}
M_{\tilde{\tau}}^2=\left(\begin{array}{ll}M_{\tilde{e}_{L_3}^{\pm} \tilde{e}_{L_3}^{\pm}}^2&M_{\tilde{e}_{L_3}^{\pm} \tilde{e}_{R_3}^{\pm}}^2\\
M_{\tilde{e}_{L_3}^{\pm} \tilde{e}_{R_3}^{\pm}}^2&M_{\tilde{e}_{R_3}^{\pm} \tilde{e}_{R_3}^{\pm}}^2\end{array}\right),
\end{eqnarray}
where the concrete expressions for matrix elements can be found in appendix~\ref{appendix-mass}. The eigenvalues $m_{\tilde{t}_{1,2}}^2$, $m_{\tilde{b}_{1,2}}^2$ and $m_{\tilde{\tau}_{1,2}}^2$ of the $\tilde{t}$, $\tilde{b}$ and $\tilde{\tau}$ mass squared matrices can be given by
\begin{eqnarray}
m_{1,2}^2={1\over 2}\Big({\rm{Tr}}{M}^2 \mp\sqrt{({{\rm{Tr}}{M}^2})^2-4{\rm{Det}}{M}^2}\Big),
\end{eqnarray}
where ${\rm{Tr}}{M}^2={M}_{11}^2+{M}_{22}^2$, ${\rm{Det}}{M}^2 = { M}_{11}^2{ M}_{22}^2-({M}_{12}^2)^2$.

Then, the mass squared matrix for the neutral Higgs doublets in the basis $(h_d,\;h_u)$ is written as
\begin{eqnarray}
&&{\cal M}^2=\left(\begin{array}{ll}M_{h_d h_d}^2&M_{h_d h_u}^2\\
M_{h_d h_u}^2&M_{h_u h_u}^2\end{array}\right) + \frac{G_F}{\sqrt{2}\pi^2} \left(\begin{array}{ll}\Delta_{11}&\Delta_{12}\\
\Delta_{12}&\Delta_{22}\end{array}\right),
\end{eqnarray}
where the dominating radiative corrections originate from fermions and corresponding supersymmetric partners in this model:
\begin{eqnarray}
\Delta_{11}=\Delta_{11}^{q}+\Delta_{11}^{l},\qquad
\Delta_{12}=\Delta_{12}^{q}+\Delta_{12}^{l},\qquad
\Delta_{22}=\Delta_{22}^{q}+\Delta_{22}^{l}.
\end{eqnarray}
Neglecting the terms containing small coupling $Y_{\nu_i}$ and $\upsilon_{\nu_i}$, and using the expressions given in refs.~\cite{loop-Higgs,Hi-1,Hi-2,Hi-3,Hi-4,Hi-5,Hi-6,Hi-7,Hi-8}, the one-loop radiative corrections from quark fields read as
\begin{eqnarray}
&&\Delta_{11}^{q}={3{m_{b}^4}\over\cos^2\beta}\Big\{\ln{m_{\tilde{b}_1}^2 m_{\tilde{b}_2}^2 \over m_{b}^4}+{2A_{b}(A_{b}-\mu\tan\beta)\over m_{\tilde{b}_1}^2-m_{\tilde{b}_2}^2} \ln{m_{\tilde{b}_1}^2\over m_{\tilde{b}_2}^2}
+{A_{b}^2(A_{b}-\mu\tan\beta)^2\over {(m_{\tilde{b}_1}^2-m_{\tilde{b}_2}^2)}^2}
g(m_{\tilde{b}_1}^2,m_{\tilde{b}_2}^2)\Big\}
\nonumber\\
&&\hspace{1.2cm}
+{3{m_{t}^4}\over\sin^2\beta}{\mu^2(A_{t}-\mu\cot\beta)^2\over {(m_{\tilde{t}_1}^2-m_{\tilde{t}_2}^2)}^2}
g(m_{\tilde{t}_1}^2,m_{\tilde{t}_2}^2),
\nonumber\\
&&\Delta_{22}^{q}={3{m_{t}^4}\over\sin^2\beta}\Big\{\ln{m_{\tilde{t}_1}^2 m_{\tilde{t}_2}^2 \over m_{t}^4}+{2A_{t}(A_{t}-\mu\cot\beta)\over m_{\tilde{t}_1}^2-m_{\tilde{t}_2}^2} \ln{m_{\tilde{t}_1}^2\over m_{\tilde{t}_2}^2}
+{A_{t}^2(A_{t}-\mu\cot\beta)^2\over {(m_{\tilde{t}_1}^2-m_{\tilde{t}_2}^2)}^2}
g(m_{\tilde{t}_1}^2,m_{\tilde{t}_2}^2)\Big\}
\nonumber\\
&&\hspace{1.2cm}
+{3{m_{b}^4}\over\cos^2\beta}{\mu^2(A_{b}-\mu\tan\beta)^2\over {(m_{\tilde{b}_1}^2-m_{\tilde{b}_2}^2)}^2}
g(m_{\tilde{b}_1}^2,m_{\tilde{b}_2}^2),
\nonumber\\
&&\Delta_{12}^{q}={3{m_{t}^4}\over\sin^2\beta}{\mu(-A_{t}+\mu\cot\beta)\over m_{\tilde{t}_1}^2-m_{\tilde{t}_2}^2}\Big\{ \ln{m_{\tilde{t}_1}^2\over m_{\tilde{t}_2}^2}
+{A_{t}(A_{t}-\mu\cot\beta)\over {(m_{\tilde{t}_1}^2-m_{\tilde{t}_2}^2)}}
g(m_{\tilde{t}_1}^2,m_{\tilde{t}_2}^2)\Big\}
\nonumber\\
&&\hspace{1.2cm}
+{3{m_{b}^4}\over\cos^2\beta}{\mu(-A_{b}+\mu\tan\beta)\over m_{\tilde{b}_1}^2-m_{\tilde{b}_2}^2} \Big\{\ln{m_{\tilde{b}_1}^2\over m_{\tilde{b}_2}^2} +{A_{b}(A_{b}-\mu\tan\beta)\over {(m_{\tilde{b}_1}^2-m_{\tilde{b}_2}^2)}}
g(m_{\tilde{b}_1}^2,m_{\tilde{b}_2}^2)\Big\},
\end{eqnarray}
with
\begin{eqnarray}
&&g(m_1^2,m_2^2)=2-{m_1^2+m_2^2\over m_1^2-m_2^2}\ln{m_1^2\over m_2^2}\;.
\end{eqnarray}
Similarly, one can obtain the one-loop radiative corrections from lepton fields
\begin{eqnarray}
&&\Delta_{11}^{l}={{m_{\tau}^4}\over\cos^2\beta}\Big\{\ln{m_{\tilde{\tau}_1}^2 m_{\tilde{\tau}_2}^2 \over m_{\tau}^4}+{2A_{\tau}(A_{\tau}-\mu\tan\beta)\over m_{\tilde{\tau}_1}^2-m_{\tilde{\tau}_2}^2} \ln{m_{\tilde{\tau}_1}^2\over m_{\tilde{\tau}_2}^2}
+{A_{\tau}^2(A_{\tau}-\mu\tan\beta)^2\over {(m_{\tilde{\tau}_1}^2-m_{\tilde{\tau}_2}^2)}^2}
g(m_{\tilde{\tau}_1}^2,m_{\tilde{\tau}_2}^2)\Big\}
\nonumber\\
&&\Delta_{22}^{l}={{m_{\tau}^4}\over\cos^2\beta}{\mu^2(A_{\tau}-\mu\tan\beta)^2\over {(m_{\tilde{\tau}_1}^2-m_{\tilde{\tau}_2}^2)}^2}
g(m_{\tilde{\tau}_1}^2,m_{\tilde{\tau}_2}^2),
\nonumber\\
&&\Delta_{12}^{l}={{m_{\tau}^4}\over\cos^2\beta}{\mu(-A_{\tau}+\mu\tan\beta)\over m_{\tilde{\tau}_1}^2-m_{\tilde{\tau}_2}^2} \Big\{\ln{m_{\tilde{\tau}_1}^2\over m_{\tilde{\tau}_2}^2} +{A_{\tau}(A_{\tau}-\mu\tan\beta)\over {(m_{\tilde{\tau}_1}^2-m_{\tilde{\tau}_2}^2)}}
g(m_{\tilde{\tau}_1}^2,m_{\tilde{\tau}_2}^2)\Big\}.
\end{eqnarray}

Then, the neutral doublet-like Higgs mass eigenvalues can be derived
\begin{eqnarray}
m_{h(H)}^2={1\over 2}\Big({\rm{Tr}}{\cal M}^2 \mp\sqrt{({{\rm{Tr}}{\cal M}^2})^2-4{\rm{Det}}{\cal M}^2}\Big).
\end{eqnarray}
One most stringent constraint on parameter space of the $\mu\nu$SSM is that the mass squared matrix should produce an eigenvalue around $(125\;{\rm GeV})^2$
as mass squared of the SM-like Higgs.
The current combination of the ATLAS and CMS data gives~\cite{mh-AC}:
\begin{eqnarray}
&&m_{{h}}=125.9\pm2.1\;{\rm GeV},
\label{M-h}
\end{eqnarray}
this fact constrains parameter space of the $\mu\nu$SSM stringently.

\section{$\mu-e$ conversion in nuclei within the $\mu\nu$SSM\label{sec:4}}
In this section, we present the analysis on the $\mu-e$ conversion processes at the quark level in the $\mu\nu$SSM. For this study we will use the indices $\beta,\zeta=1,\ldots,5$, $I=1,\ldots,6$, $\alpha,\rho=1,\ldots,8$, and $\eta,\sigma=1,\ldots,10$. The summation convention is implied on the repeated indices in the following.

\begin{figure}[htbp]
\setlength{\unitlength}{1mm}
\centering
\includegraphics[width=4.6in]{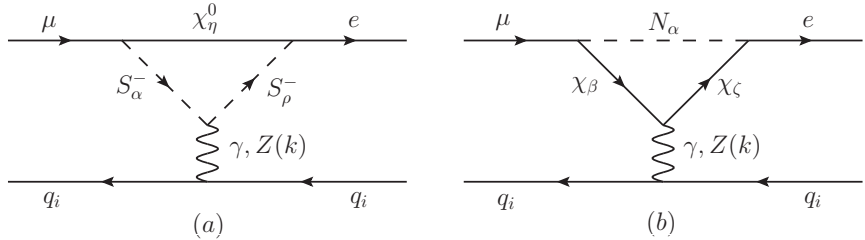}
\caption[]{Penguin-type diagrams for the $\mu-e$ conversion processes at the quark level. (a) represents the contributions from neutral fermions $\chi_\eta^0$ and charged scalars $S_\alpha^-$ loops, and (b) represents the contributions from charged fermions $\chi_\beta$ and neutral scalars $N_\alpha$ ($N=S,P$) loops.}
\label{fig1}
\end{figure}

Figure~\ref{fig1} shows the penguin-type diagrams for the $\mu-e$ conversion processes at the quark level  in the $\mu\nu$SSM. We will give the effective Lagrangian of the process at the quark level. The $\gamma$-penguin-type diagrams give the terms~\cite{Hisano}
\begin{eqnarray}
&&\mathcal{L}_{int}^{\gamma - {\rm{p}}} = -\frac{{{e^2}}}{{{k^2}}} {\bar e}\Big[{k^2}{\gamma _\alpha }(A_1^L{P_L}+ A_1^R{P_R}) + {m_{\mu}}i{\sigma _{\alpha \beta}}{k^\beta }(A_2^L{P_L}
+ A_2^R{P_R})\Big] \mu  \nonumber\\
&&\qquad\quad\;\; \times \:\sum\limits_{q=u,d} Q_{em}^q{\bar q}{\gamma ^\alpha }q.
\end{eqnarray}
where  $P_L=\frac{1}{2}{(1 - {\gamma _5})}$, $P_R=\frac{1}{2}{(1 + {\gamma _5})}$, $Q_{em}^u=\frac{2}{3}$, $Q_{em}^d=-\frac{1}{3}$ and ${m_{\mu}}$ is the muon mass, respectively.  Each coefficient in the above can be written as
\begin{eqnarray}
A_a^{L,R} = A_a^{(n)L,R} + A_a^{(c)L,R} \quad (a = 1,2),
\end{eqnarray}
where $A_a^{(n)L,R}$ denote for the contributions from the virtual neutral fermion loops, and $A_a^{(c)L,R}$ denote that from the virtual charged fermion loops, respectively. After integrating the heavy freedoms out, we formulate those coefficients as follows
\begin{eqnarray}
&&A_1^{(n)L} = \frac{1}{6{m_W^2}}C_R^{S_\alpha^- \chi_\eta^\circ {{\bar \chi }_3}}C_L^{S_\alpha^{-\ast} {\chi _4}\bar \chi _\eta ^ \circ }{I_4}({x_{\chi _\eta ^ \circ }},{x_{S_\alpha ^ -}}),\nonumber\\
&&A_2^{(n)L} = \frac{{{m_{\chi _\eta ^ \circ }}}}{{m_{\mu}}{m_W^2}}C_L^{S_\alpha ^ - \chi _\eta ^ \circ {{\bar \chi }_3}}C_L^{S_\alpha ^{-\ast} {\chi _4}\bar \chi _\eta ^ \circ }
\Big[ {I_3}({x_{\chi _\eta ^ \circ }},{x_{S_\alpha ^ - }}) - {I_1}(
{x_{\chi _\eta ^ \circ }},{x_{S_\alpha ^ - }}) \Big], \nonumber\\
&&A_a^{(n)R} = \left. {A_a^{(n)L}} \right|{ _{L \leftrightarrow R}},
\end{eqnarray}
where the concrete expressions for form factors $I_k \;(k=1,\ldots,4)$ can be found in appendix~\ref{appendix-integral}.
Additionally, $x_i= {m_i^2}/{m_W^2}$ and $m_i$ is the mass for the corresponding particle. In a similar way, the corrections from the Feynman diagrams with virtual charged fermions are
\begin{eqnarray}
&&A_1^{(c)L} = \sum\limits_{N=S,P} \frac{1}{6{m_W^2}}
C_R^{{N_\alpha }{\chi _\beta }{{\bar \chi }_3}}C_L^{{N_\alpha }{\chi _4}{{\bar \chi }_\beta }}
\Big[ {I_1}({x_{{\chi _\beta }}},{x_{{N_\alpha }}})- 2 {I_2}({x_{{\chi _\beta }}},{x_{{N_\alpha }}}) - {I_4}({x_{{\chi _\beta }}},{x_{{N_\alpha }}}) \Big] , \nonumber\\
&&A_2^{(c)L} = \sum\limits_{N=S,P} \frac{{{m_{{\chi _{^\beta }}}}}}{{m_{\mu}}{m_W^2}}
C_L^{{N_\alpha }{\chi _\beta }{{\bar \chi }_3}}
C_L^{{N_\alpha }{\chi _4}{{\bar \chi }_\beta }}\Big[ {I_1}(
{x_{{\chi _\beta }}},{x_{{N_\alpha }}}) - {I_2}({x_{{\chi _\beta }}},{x_{{N_\alpha }}}) - {I_4}({x_{{\chi _\beta }}},{x_{{N_\alpha }}}) \Big] , \nonumber\\
&&A_a^{(c)R} = \left. {A_a^{(c)L}} \right|{ _{L \leftrightarrow R}}.
\end{eqnarray}

Similarly, the effective Lagrangian of $Z$-penguin-type diagrams is
\begin{eqnarray}
&&\mathcal{L}_{int}^{Z - {\rm{p}}} = \frac{{{e^2}}}{{m_Z^2}s_{_W}^2c_{_W}^2}\sum\limits_{q=u,d}\frac{Z_L^q+Z_R^q}{2}{\bar q}\gamma _\alpha q {\bar e}{\gamma ^\alpha }({F_L}{P_L} + {F_R}{P_R})\mu  ,
\end{eqnarray}
where
\begin{eqnarray}
Z_{L,R}^q = T_{3L,R}^q - Q_{em}^q s_{_W}^2,\quad (q=u,d),
\end{eqnarray}
with $T_{3L}^u=\frac{1}{2}$, $T_{3L}^d=-\frac{1}{2}$ and $T_{3R}^u=T_{3R}^d=0$. And the coefficient $F_{L,R}$ also can be written as
\begin{eqnarray}
{F_{L,R}} = F_{L,R}^{(n)} + F_{L,R}^{(c)},
\end{eqnarray}
where the contributions to the effective couplings $F_{L,R}^{(n)}$ and $F_{L,R}^{(c)}$ are
\begin{eqnarray}
&&F_L^{(n)} = \sum\limits_{N=S,P} \Big[ \frac{{m_{{\chi _\zeta }}}{m_{{\chi _\beta }}}}
{{e^2}{m_W^2}}C_R^{{N_\alpha }{\chi _\zeta }{{\bar \chi }_3}}C_L^{Z{\chi _\beta }{{\bar \chi }_\zeta }}
C_L^{{N_\alpha }{\chi _4}{{\bar \chi }_\beta }}{G_1}({x_{{N_\alpha }}},{x_{{\chi _\zeta }}},{x_{{\chi _\beta }}})\nonumber\\
&&\qquad\;\;\quad - \: \frac{1}{2{e^2}} C_R^{{N_\alpha }{\chi _\zeta }{{\bar \chi }_3}}C_R^{Z{\chi _\beta }
{{\bar \chi }_\zeta }}C_L^{{N_\alpha }{\chi _4}{{\bar \chi }_\beta }}{G_2}({x_{{N_\alpha }}},
{x_{{\chi _\zeta }}},{x_{{\chi _\beta }}}) \Big],
\nonumber\\
&&F_L^{(c)} = \, \frac{1}{2{e^2}}C_R^{S_\rho ^ - \chi _\eta ^0{{\bar \chi }_3}}
C_R^{ZS_\alpha ^ - S_\rho ^ {-\ast} }C_L^{S_\alpha ^{-\ast} {\chi _4}\bar \chi _\eta ^0}{G_2}
({x_{\chi _\eta ^0}},{x_{S_\alpha ^ - }},{x_{S_\rho ^ - }}) ,
\nonumber\\
&&F_R^{(n,c)} = \left. {F_L^{(n,c)}} \right|{ _{L \leftrightarrow R}} .
\end{eqnarray}
Here, the concrete expressions for $G_k\;(k=1,\ldots,4)$ can be found in appendix~\ref{appendix-integral}.

\begin{figure}[htbp]
\setlength{\unitlength}{1mm}
\centering
\begin{minipage}[c]{0.7\textwidth}
\includegraphics[width=4.6in]{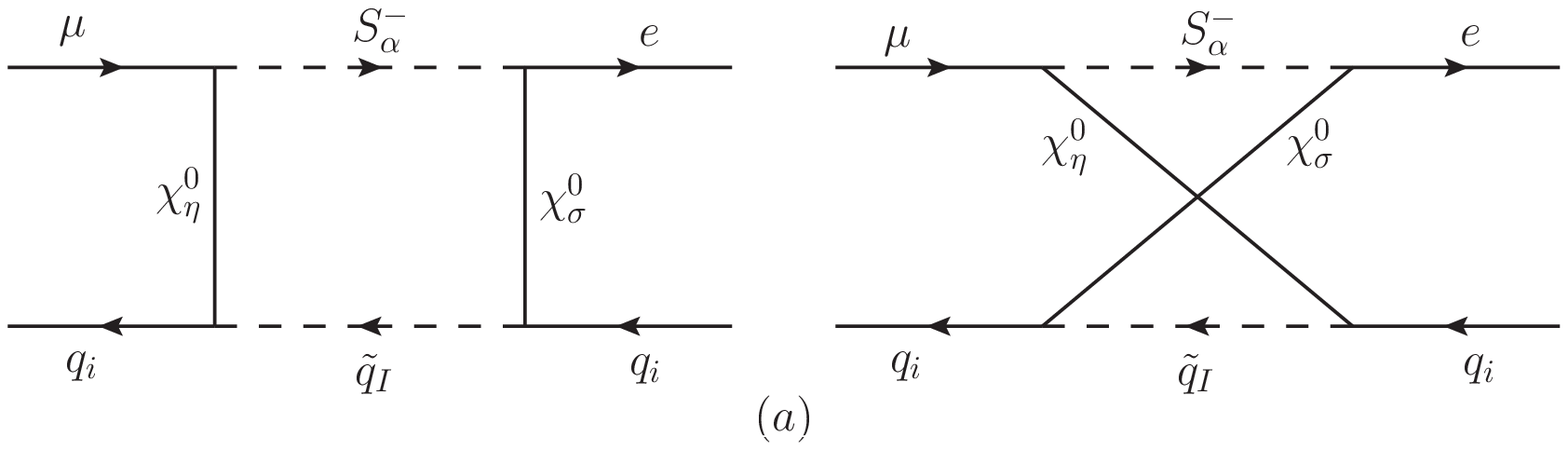}
\label{fig2a}
\end{minipage}
\begin{minipage}[c]{0.7\textwidth}
\includegraphics[width=4.6in]{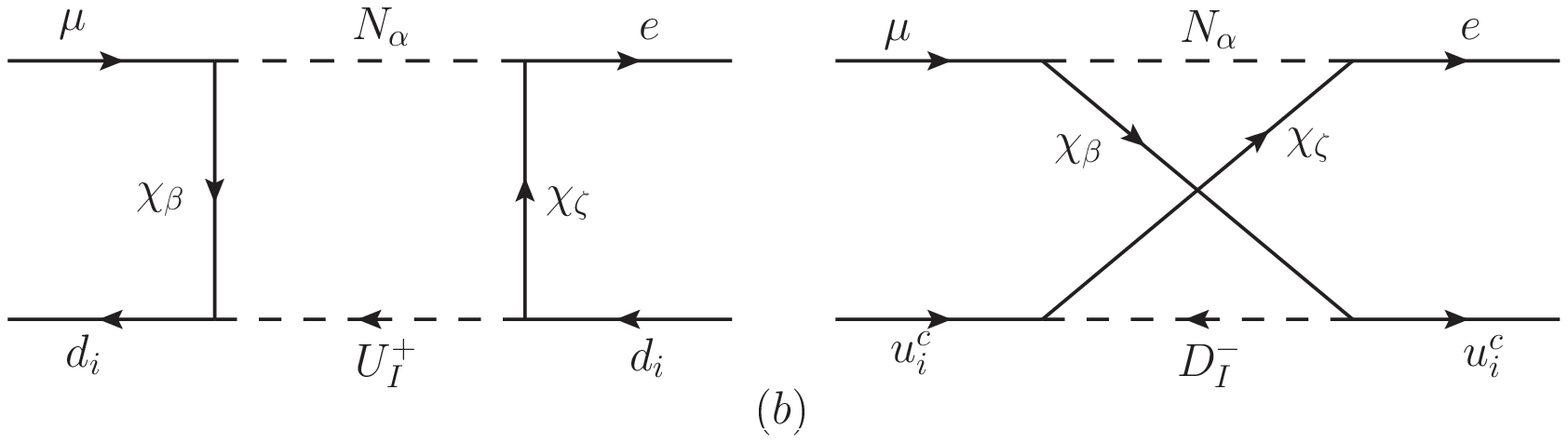}
\label{fig2b}
\end{minipage}
\caption[]{Box-type diagrams for the $\mu-e$ conversion processes at the quark level. (a) represents the contributions from neutral fermions $\chi_\eta^0$, charged scalars $S_\alpha^-$ and squark $\tilde{q}_I$ ($q=u,d$ and $\tilde{u}_I=U_I^+$, $\tilde{d}_I=D_I^-$) loops, and (b) represents the contributions from charged fermions $\chi_\beta$ and neutral scalars $N_\alpha$ ($N=S,P$) and squark $\tilde{q}_I$  loops.}
\label{fig2}
\end{figure}

Furthermore, the effective Lagrangian from the box-type diagrams drawn in figure~\ref{fig2} can be written as
\begin{eqnarray}
&&\mathcal{L}_{int}^{box}  = e^2 \sum\limits_{q=u,d}{\bar q}\gamma _\alpha q {\bar e}{\gamma ^\alpha }({D_q^L}{P_L} + {D_q^R}{P_R})\mu ,
\end{eqnarray}
with
\begin{eqnarray}
D_q^{L,R} = D_q^{(n)L,R} + D_q^{(c)L,R} \quad (q=u,d).
\end{eqnarray}
The effective couplings $D_q^{(n)L,R}$ originate from those box diagrams with virtual neutral fermion contributions:
\begin{eqnarray}
&&D_q^{(n)L} = \frac{1}{8{e^2}{m_W^2}}{G_4}({x_{\chi_\eta^0}},{x_{\chi_\sigma^0}},{x_{S_\alpha^-}},
{x_{\tilde{q}_I}})\Big[
C_R^{S_\alpha ^ - \chi _\eta^0{{\bar \chi }_4}\ast} C_R^{S_\alpha^ - \chi _\sigma^0{{\bar \chi }_3}}
C_R^{{\tilde{q}_I} \chi _\eta^0{{\bar q }_i}}C_R^{{\tilde{q}_I} \chi _\sigma^0{{\bar q }_i}\ast}
\nonumber\\
&&\qquad\qquad - \:C_R^{S_\alpha ^ - \chi _\eta^0{{\bar \chi }_4}\ast} C_R^{S_\alpha^ - \chi _\sigma^0{{\bar \chi }_3}}C_L^{{\tilde{q}_I} \chi _\eta^0{{\bar q }_i}\ast}C_L^{{\tilde{q}_I} \chi _\sigma^0{{\bar q }_i}}\Big] \nonumber\\
&&\qquad\qquad - \:\frac{{m_{\chi_\eta^0}}{m_{\chi_\sigma^0}}}{4{e^2}{m_W^4}}
{G_3}({x_{\chi_\eta^0}},{x_{\chi_\sigma^0}},{x_{S_\alpha^-}},
{x_{\tilde{q}_I}})\Big[
C_R^{S_\alpha ^ - \chi _\eta^0{{\bar \chi }_4}\ast} C_R^{S_\alpha^ - \chi _\sigma^0{{\bar \chi }_3}}
C_L^{{\tilde{q}_I} \chi _\eta^0{{\bar q }_i}}C_L^{{\tilde{q}_I} \chi _\sigma^0{{\bar q }_i}\ast}
\nonumber\\
&&\qquad\qquad - \:C_R^{S_\alpha ^ - \chi _\eta^0{{\bar \chi }_4}\ast} C_R^{S_\alpha^ - \chi _\sigma^0{{\bar \chi }_3}}C_R^{{\tilde{q}_I} \chi _\eta^0{{\bar q }_i}\ast}C_R^{{\tilde{q}_I} \chi _\sigma^0{{\bar q }_i}}\Big] ,
\nonumber\\
&&D_q^{(n)R} = \left. {D_q^{(n)L}} \right|
{ _{L \leftrightarrow R}} \:,\:\;(q=u,d \;\textrm{and}\; \tilde{u}_I=U_I^+, \:\tilde{d}_I=D_I^-).
\end{eqnarray}
Correspondingly, the effective couplings from the box diagrams with virtual charged fermion contributions $D_q^{(c)L,R}$ are
\begin{eqnarray}
&&D_d^{(c)L} = \sum\limits_{N=S,P} \Big[ \frac{1}{8{e^2}{m_W^2}} {G_4}({x_{{\chi _\beta}}},{x_{{\chi _\zeta }}}
,{x_{{N_\alpha}}},{x_{U_I^+}})C_R^{{N_\alpha }{\chi _\beta}{{\bar \chi }_4}\ast}
C_R^{{N_\alpha}{\chi _\zeta}{{\bar \chi }_3}}C_R^{U_I^+ {\chi _\beta }{{\bar d}_i}}
C_R^{U_I^+ {\chi _\zeta }{{\bar d}_i}\ast}\nonumber\\
&&\qquad\qquad - \: \frac{{m_{\chi _\beta}}{m_{\chi _\zeta}}}{4{e^2}{m_W^4}}
{G_3}({x_{{\chi _\beta}}},{x_{{\chi _\zeta }}}
,{x_{{N_\alpha}}},{x_{U_I^+}})C_R^{{N_\alpha }{\chi _\beta}{{\bar \chi }_4}\ast}
C_R^{{N_\alpha}{\chi _\zeta}{{\bar \chi }_3}}C_L^{U_I^+ {\chi _\beta }{{\bar d}_i}}
C_L^{U_I^+ {\chi _\zeta }{{\bar d}_i}\ast} \Big] ,
\nonumber\\
&&D_u^{(c)L} = \sum\limits_{N=S,P} \Big[ -\frac{1}{8{e^2}{m_W^2}} {G_4}({x_{{\chi _\beta}}},{x_{{\chi _\zeta }}}
,{x_{{N_\alpha}}},{x_{D_I^-}})C_R^{{N_\alpha }{\chi _\beta}{{\bar \chi }_4}\ast}
C_R^{{N_\alpha}{\chi _\zeta}{{\bar \chi }_3}}C_L^{D_I^- {u_i^c} {\bar \chi _\beta }\ast}
C_L^{D_I^- {u_i^c} {\bar \chi _\zeta }}\nonumber\\
&&\qquad\qquad + \: \frac{{m_{\chi _\beta}}{m_{\chi _\zeta}}}{4{e^2}{m_W^4}}
{G_3}({x_{{\chi _\beta}}},{x_{{\chi _\zeta }}}
,{x_{{N_\alpha}}},{x_{D_I^-}})C_R^{{N_\alpha }{\chi _\beta}{{\bar \chi }_4}\ast}
C_R^{{N_\alpha}{\chi _\zeta}{{\bar \chi }_3}}C_R^{D_I^- {u_i^c} {\bar \chi _\beta }\ast}
C_R^{D_I^- {u_i^c} {\bar \chi _\zeta }} \Big] ,
\nonumber\\
&&D_q^{(c)R} = \left. {D_q^{(c)L}} \right|
{ _{L \leftrightarrow R}} ,\quad (q=u,d ).
\end{eqnarray}

Using the expression for the effective Lagrangian of the $\mu-e$ conversion processes at the quark level, we can calculate the $\mu-e$ conversion rate in a nucleus~\cite{Bernabeu}:
\begin{eqnarray}
&&{\rm{CR}}(\mu \to e:{\rm{Nucleus}}) \nonumber\\
&&\qquad = 4 \alpha^5 \frac{Z_{\rm{eff}}^4}{Z } \left| F(q^2) \right|^2 m_\mu^5  \Big[\left| Z( A_1^L  -  {A_2^R} ) - (2Z+N)\bar{D}_u^L - (Z+2N)\bar{D}_d^L \right| ^2 \nonumber\\
&&\qquad\quad + \: \left| Z( A_1^R  -  {A_2^L} ) - (2Z+N)\bar{D}_u^R - (Z+2N)\bar{D}_d^R \right|^2 \Big]\frac{1}{\Gamma_{\rm{capt}}},
\label{gamma-2}
\end{eqnarray}
with
\begin{eqnarray}
&&\bar{D}_q^L = D_q^L + \frac{Z_L^q+Z_R^q}{2} \frac{F_L}{{m_Z^2}s_{_W}^2c_{_W}^2} , \nonumber\\
&&\bar{D}_q^R  =  \left. {\bar{D}_q^L} \right|{ _{L \leftrightarrow R}}  \quad (q=u,d ), \quad
\end{eqnarray}
where $Z$ and $N$ denote the proton and neutron numbers in a nucleus, while $Z_{\rm{eff}}$ is an effective atomic charge which has been determined in refs.~\cite{Zeff,Zeff1}. $F(q^2)$ is the nuclear form factor and $\Gamma_{\rm{capt}}$ denotes the total muon capture rate. The values of $Z_{\rm{eff}}$, $F(q^2\simeq-m_\mu^2)$ and $\Gamma_{\rm{capt}}$ for different nuclei have been collect in table~\ref{tab1} and follow ref.~\cite{Kitano}.
\begin{table*}
\begin{tabular*}{\textwidth}{@{\extracolsep{\fill}}lllll@{}}
\hline
$_{Z}^{A}{\rm{Nucleus}}$ & \multicolumn{1}{c}{$Z_{\rm{eff}}$} & \multicolumn{1}{c}{$F(q^2\simeq-m_\mu^2)$} & $\Gamma_{\rm{capt}}({\rm{GeV}})$ \\
\hline
$_{22}^{48}{\rm{Ti}}$ & 17.6 & 0.54 & $1.70422\times10^{-18}$ \\
$_{\:79}^{197}{\rm{Au}}$ & 33.5 & 0.16 & $8.59868\times10^{-18}$ \\
$_{\:82}^{207}{\rm{Pb}}$ & 34.0 & 0.15 & $8.84868\times10^{-18}$ \\
\hline
\end{tabular*}
\caption{The values of $Z_{\rm{eff}}$, $F(q^2\simeq-m_\mu^2)$ and $\Gamma_{(\rm{capt})}$ for different nuclei}
\label{tab1}
\end{table*}

In the last, we also can obtain the branching ratio for $\mu\rightarrow e\gamma$ as
\begin{eqnarray}
{\rm{Br}} (\mu\rightarrow e\gamma) = \frac{{{e^2}}}{{16\pi }}m_\mu^5 \Big({\left| {A_2^L} \right|^2} + {\left| {A_2^R} \right|^2}\Big)\frac{1}{\Gamma_\mu},
\label{gamma-1}
\end{eqnarray}
where $\Gamma_\mu \approx 2.996 \times 10^{-19}\:{\rm{GeV}}$ denotes the total decay rate of the muon.

\section{The numerical results\label{sec:5}}

\subsection{The parameter space}
It is well known that there are many free parameters in various SUSY extensions of the SM.
In order to obtain a more transparent numerical results, we take some assumptions on parameter space of the $\mu\nu{\rm SSM}$ before we perform the numerical analysis. We adopt the minimal flavor violation (MFV) assumptions
\begin{eqnarray}
&&\;\,{\kappa _{ijk}} = \kappa {\delta _{ij}}{\delta _{jk}}, \;\;
{({A_\kappa }\kappa )_{ijk}} =
{A_\kappa }\kappa {\delta _{ij}}{\delta _{jk}}, \quad\;
\lambda _i = \lambda , \qquad\quad\;\;
{{\rm{(}}{A_\lambda }\lambda {\rm{)}}_i} = {A_\lambda }\lambda,\nonumber\\
&&\;\,{Y_{{u _{ij}}}} = {Y_{{u _i}}}{\delta _{ij}},\quad
 (A_u Y_u)_{ij}={A_u}{Y_{{u_i}}}{\delta _{ij}},\quad
{Y_{{\nu _{ij}}}} = {Y_{{\nu _i}}}{\delta _{ij}},\quad (A_\nu Y_\nu)_{ij}={A_\nu}{Y_{{\nu_i}}}{\delta _{ij}},\nonumber\\
&&\;\,{Y_{{d_{ij}}}} = {Y_{{d_i}}}{\delta _{ij}},\quad\:
(A_d Y_d)_{ij}={A_d}{Y_{{d_i}}}{\delta _{ij}},\quad\,
{Y_{{e_{ij}}}} = {Y_{{e_i}}}{\delta _{ij}},\quad\,
{({A_e}{Y_e})_{ij}} = {A_e}{Y_{{e_i}}}{\delta _{ij}},
\nonumber\\
&&m_{{{\tilde L}_{ij}}}^2 = m_{\tilde L_i}^2{\delta _{ij}},\qquad\;\;
m_{\tilde \nu_{ij}^c}^2 = m_{{{\tilde \nu_i}^c}}^2{\delta _{ij}},\quad\;\;\:
m_{\tilde e_{ij}^c}^2 = m_{{{\tilde e}^c}}^2{\delta _{ij}},\qquad\quad \upsilon_{\nu_i^c}=\upsilon_{\nu^c}, \nonumber\\
&&m_{\tilde Q_{ij}}^2 = m_{{{\tilde Q}}}^2{\delta _{ij}}, \qquad\;\;
m_{\tilde u_{ij}^c}^2 = m_{{{\tilde u}^c}}^2{\delta _{ij}}, \quad\;\;\:
m_{\tilde d_{ij}^c}^2 = m_{{{\tilde d}^c}}^2{\delta _{ij}},
\label{MFV}
\end{eqnarray}
where $i,\;j,\;k =1,\;2,\;3 $. Then, we can have ${A_t}\equiv{A_u}$, ${A_b}\equiv{A_d}$ and ${A_\tau}\equiv{A_e}$.

Restrained by the quark and lepton masses, we have
\begin{eqnarray}
{Y_{{u_i}}} = \frac{{{m_{{u_i}}}}}{{{\upsilon_u}}},\qquad {Y_{{d_i}}} = \frac{{{m_{{d_i}}}}}{{{\upsilon_d}}},\qquad {Y_{{e_i}}} = \frac{{{m_{{l_i}}}}}{{{\upsilon_d}}},
\end{eqnarray}
where $m_{u_i}$, $m_{d_i}$ and $m_{l_i}$ are the up-quark, down-quark and charged lepton masses, respectively. Here, we choose the values of the fermion masses from ref.~\cite{PDG}. At the EW scale, the soft masses $m_{\tilde H_d}^2$, $m_{\tilde H_u}^2$, $m_{\tilde L_i}^2$ and $m_{\tilde \nu_i^c}^2$ can be derived from the minimization conditions of the tree-level neutral scalar potential, which are given in appendix~\ref{appendix-mini}.

The $3\times3$ matrix $Y_\nu$ determines the Dirac masses for the neutrinos ${Y_\nu}{\upsilon_u} \sim {m_D}$, and the tiny neutrino masses are obtained through TeV scale seesaw mechanism $m_\nu\sim m_D m_{_N}^{-1}m_D^T$. This indicates that the nonzero VEVs of left-handed sneutrinos satisfy $\upsilon_{\nu_i}\ll\upsilon_{u,d}$, then
\begin{eqnarray}
\tan\beta\simeq \frac{\upsilon_u}{\upsilon_d}.
\end{eqnarray}
Assuming that the charged lepton mass matrix in the flavor basis is in the diagonal form, we parameterize the unitary matrix which diagonalizes the effective light neutrino mass matrix $m_{eff}$ (can be found in appendix~\ref{appendix-approximate}) as~\cite{Uv1,Uv2,Uv3}
\begin{eqnarray}
{U_\nu} = &&\left( {\begin{array}{*{20}{c}}
   {{c_{12}}{c_{13}}} & {{s_{12}}{c_{13}}} & {{s_{13}}{e^{ - i\delta }}}  \\
   { - {s_{12}}{c_{23}} - {c_{12}}{s_{23}}{s_{13}}{e^{i\delta }}} & {{c_{12}}{c_{23}} - {s_{12}}
   {s_{23}}{s_{13}}{e^{i\delta }}} & {{s_{23}}{c_{13}}}  \\
   {{s_{12}}{s_{23}} - {c_{12}}{c_{23}}{s_{13}}{e^{i\delta }}} & { - {c_{12}}{s_{23}} - {s_{12}}
   {c_{23}}{s_{13}}{e^{i\delta }}} & {{c_{23}}{c_{13}}}  \\
\end{array}} \right)\nonumber\\
&&\: \times \: diag(1,{e^{i\frac{{{\alpha _{21}}}}{2}}},{e^{i\frac{{{\alpha _{31}}}}{2}}}),
\label{PMNS-matrix}
\end{eqnarray}
where ${c_{ij}} = \cos {\theta _{ij}}$, ${s_{ij}} = \sin {\theta _{ij}}$, the angles
${\theta _{ij}} = \left[\: {0,\pi/2} \:\right]$, $\delta = \left[ \:{0,2\pi } \:\right]$
is the Dirac CP violation phase and $\alpha_{21}$, $\alpha_{31}$ are two Majorana CP violation phases, respectively. Here,  we choose $\delta=\alpha_{21}=\alpha_{31}=0$.
$U_\nu$ diagonalizes $m_{eff}$ in the following way:
\begin{eqnarray}
U_\nu ^T m_{eff}^T{m_{eff}}{U_\nu} = diag({m_{\nu _1}^2},{m_{\nu _2}^2},{m_{\nu _3}^2}),
\label{neutrino-diagonalize}
\end{eqnarray}
where the neutrino masses $m_{\nu _i}$ connected with experimental measurements through
\begin{eqnarray}
 {m_{\nu_2}^2 - m_{\nu_1}^2}  =  {\Delta m_{21}^2} , \qquad
 {m_{\nu_3}^2 - m_{\nu_2}^2}  =  {\Delta m_{32}^2} .
\label{mass-squared-neutrino}
\end{eqnarray}
The combination of eqs.~\eqref{PMNS-matrix}, \eqref{neutrino-diagonalize} and \eqref{mass-squared-neutrino} with neutrino oscillation experimental data gives  constraint on relevant parameter space of the $\mu\nu$SSM.

Concerning the rest of the soft parameters, we will take for simplicity in the following computation $m_{{\tilde Q},{\tilde u}^c,{\tilde d}^c,{\tilde e}^c}=800\:{\rm{GeV}}$, $A_{u,d,e}=500\:{\rm{GeV}}$ and the approximate GUT relation $M_1=\frac{\alpha_1^2}{\alpha_2^2}M_2\approx 0.5 M_2$. Then, the free parameters affect our analysis are
\begin{eqnarray}
\lambda , \: \kappa ,\: \tan \beta , \: {A_\lambda },\: {A_\kappa }, \: {A_\nu }, \: {\upsilon_{\nu^c}}, \:M_2.
\end{eqnarray}

To obtain the Yukawa couplings $Y_{\nu_i}$ and $\upsilon_{\nu_i}$ from eq.~\eqref{neutrino-diagonalize},
we assume the neutrinos masses satisfying  normal hierarchy  ${m_{\nu_1}}{\rm{ < }}{m_{\nu_2}}{\rm{ < }}{m_{\nu_3}}$,
and choose $m_{\nu_2}=10^{-2}\:{\rm{eV}}$ as input in our numerical analysis. Then we can get $m_{\nu_{1,3}}$ from the experimental data on the differences of neutrino mass squared. For $U_\nu$, the values of $\theta_{ij}$ are obtained from the experimental data presented in eqs.~\eqref{neutrino-oscillations1} and \eqref{neutrino-oscillations2}. And the effective light neutrino mass matrix $m_{eff}$ can approximate as~\cite{neutrino-mass}
\begin{eqnarray}
{m_{ef{f_{ij}}}} \approx \frac{{2A{\upsilon_{\nu^c}}}}{{3\Delta }}{b_i}{b_j} + \frac{{1 - 3{\delta _{ij}}}}{{6\kappa {\upsilon_{\nu^c}}}}{a_i}{a_j},
\end{eqnarray}
where
\begin{eqnarray}
&&\Delta  = {\lambda ^2}{(\upsilon_d^2 + \upsilon_u^2)}^2 + 4\lambda \kappa {\upsilon_{\nu^c}^2}{\upsilon_d}{\upsilon_u} - 12{\lambda ^2}{\upsilon_{\nu^c}}AB,\nonumber\\
&& A = \kappa {\upsilon_{\nu^c}^2} + \lambda {\upsilon_d}{\upsilon_u},\quad \frac{1}{B}= \frac{e^2}{c_{_W}^2{M_1}} + \frac{e^2}{s_{_W}^2{M_2}} , \nonumber\\
&&{a_i} = {Y_{{\nu _i}}}{\upsilon_u}\:, \qquad\qquad\;\: {b_i} = {Y_{{\nu _i}}}{\upsilon_d} + 3\lambda \upsilon_{\nu_i}.
\end{eqnarray}
Then, we can numerically derive $Y_{\nu_i} \sim \mathcal{O}(10^{-7})$ and $\upsilon_{\nu_i} \sim \mathcal{O}(10^{-4}{\rm{GeV}})$ from eq.~\eqref{neutrino-diagonalize}.

\subsection{$\mu-e$ conversion rates in nuclei with a 125 GeV Higgs}

\begin{figure}[htbp]
\setlength{\unitlength}{1mm}
\centering
\begin{minipage}[c]{0.7\textwidth}
\includegraphics[width=3.3in]{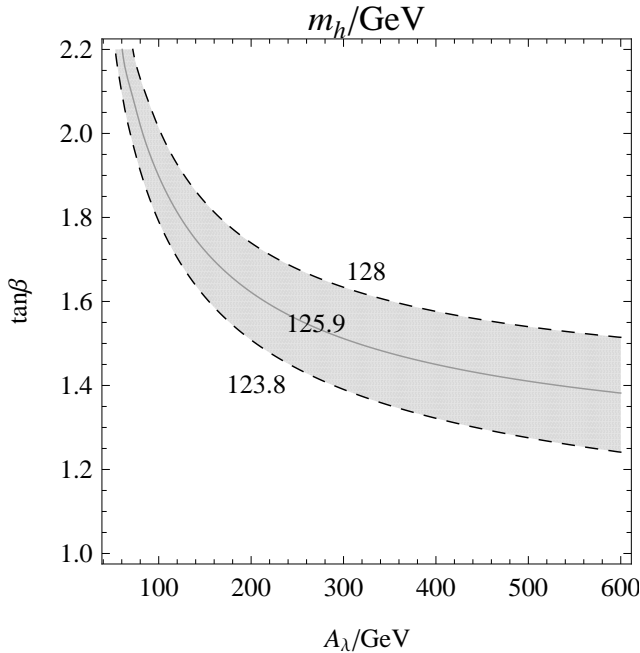}
\end{minipage}
\caption[]{Contour plot of the SM-like Higgs mass in the ($A_\lambda$-$\tan \beta$) plane.}
\label{fig-mh}
\end{figure}

Considering the research of the $\mu\nu{\rm SSM}$~\cite{mnSSM1,mnSSM2,Zhang}, we choose the relevant parameters as $\lambda=0.1$, $\kappa=0.01$, $A_\nu=A_\kappa=-1\:{\rm{TeV}}$, $M_2=3\:{\rm{TeV}}$ and $\upsilon_{\nu^c}=800\:{\rm{GeV}}$ in next numerical analysis for convenience. With those assumptions on parameter space, we show the contour plot of the SM-like Higgs mass in the ($A_\lambda$-$\tan \beta$) plane in figure~\ref{fig-mh}. Constrained by the ATLAS and CMS data in eq.~\eqref{M-h}, the result indicate that the experimental data favor small $\tan \beta$. Along with increasing of $\tan \beta$, the SM-like Higgs mass increases rapidly. As $\tan \beta =1.4$, the SM-like Higgs mass increases gently with increasing of $A_\lambda$. Ensured that the SUSY partner masses are large, the value of $A_\lambda$ need to be large. In next numerical calculation, we choose $A_\lambda=500$ GeV and $\tan \beta =1.4$, keeping the SM-like Higgs mass around 125 GeV.

\begin{figure}[htbp]
\setlength{\unitlength}{1mm}
\centering
\begin{minipage}[c]{0.7\textwidth}
\includegraphics[width=3.5in]{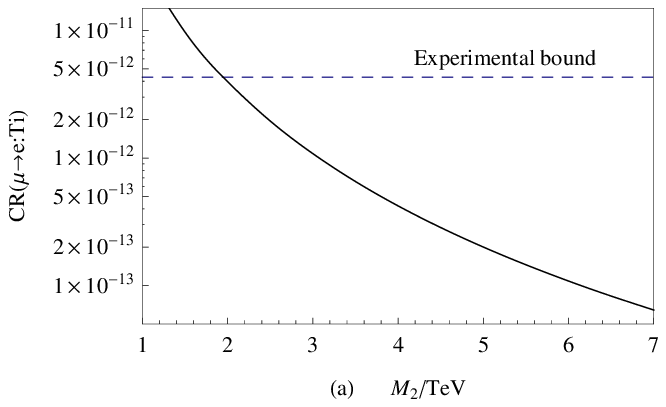}
\end{minipage}
\begin{minipage}[c]{0.7\textwidth}
\includegraphics[width=3.5in]{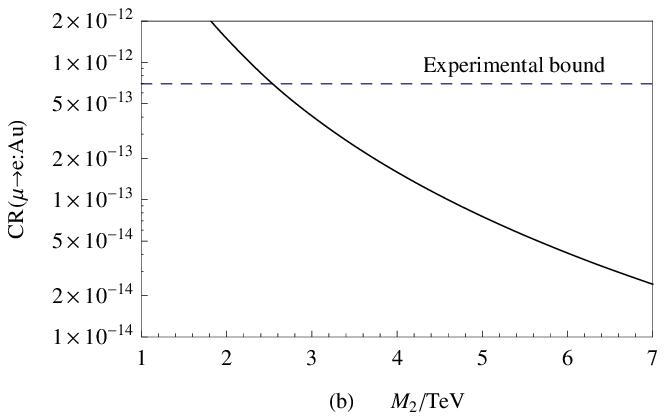}
\end{minipage}
\begin{minipage}[c]{0.7\textwidth}
\includegraphics[width=3.5in]{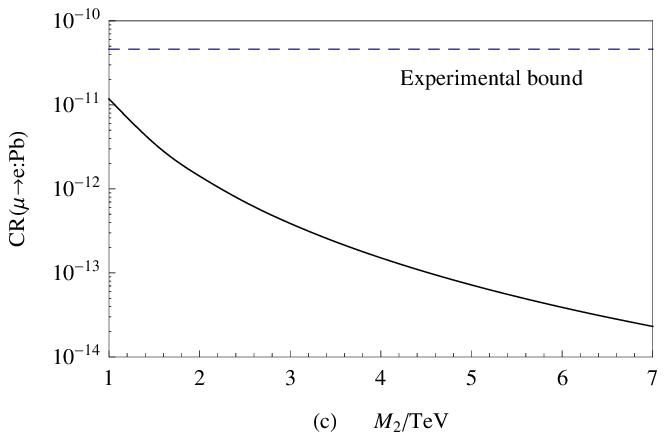}
\end{minipage}
\caption[]{The $\mu-e$ conversion rates in nuclei vary with $M_2$ for (a)  $_{22}^{48}{\rm{Ti}}$, (b) $_{\:79}^{197}{\rm{Au}}$, and (c)  $_{\:82}^{207}{\rm{Pb}}$.}
\label{fig-CR}
\end{figure}

With above assumptions on parameter space, we investigate the $\mu-e$  conversion in nuclei within the $\mu\nu{\rm SSM}$ in detail. Due to the value of $M_2$ is inert for the SM-like Higgs mass, we present the $\mu-e$ conversion rates vary with $M_2$ for different nuclei $_{22}^{48}\rm{Ti}$, $_{\:79}^{197}\rm{Au}$ and $_{\:82}^{207}\rm{Pb}$ in figure~\ref{fig-CR}, which show that the $\mu-e$ conversion rates in nuclei decrease with increasing of $M_2$. In the figure~\ref{fig-CR}(a) and figure~\ref{fig-CR}(b), we can see that the $\mu-e$ conversion rates in nuclei $_{22}^{48}\rm{Ti}$ and $_{\:79}^{197}\rm{Au}$ exceed the upper experimental bound easily, when $M_2\leq 2$TeV. The fact implies that experimental data do not favor small $M_2$.  By Introducing the left- and right-handed sneutrinos which the VEVs are nonzero to the $\mu\nu$SSM, the $\mu-e$ conversion rates in nuclei $_{22}^{48}\rm{Ti}$ and $_{\:79}^{197}\rm{Au}$  can easily reach the upper experimental bound. In the future experiments, one expects significant improvements in the sensitivities to the $\mu-e$ conversion rates in nuclei. For example, the DeeMe experiment aims at reaching a sensitivity of $10^{-14}$ level by 2015~\cite{DeeMe}. Undoubtedly, the COMET at J-PARK~\cite{COMET1,COMET2} and Mu2e at Fermilab~\cite{Mu2e1,Mu2e2} would provide the potentially measurable sensitivity of $\mathcal{O}(10^{-17})$ in the near future. With the future experimental sensitivities, the $\mu-e$ conversion in nuclei within the $\mu\nu$SSM could be detected.

\begin{figure}[htbp]
\setlength{\unitlength}{1mm}
\centering
\begin{minipage}[c]{0.7\textwidth}
\includegraphics[width=3.9in]{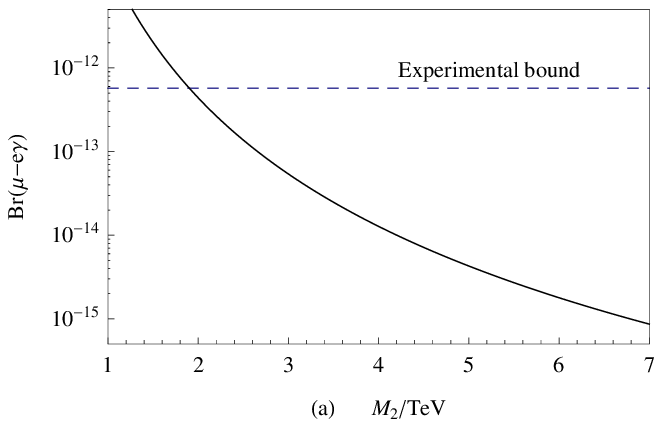}
\end{minipage}
\begin{minipage}[c]{0.7\textwidth}
\includegraphics[width=3.9in]{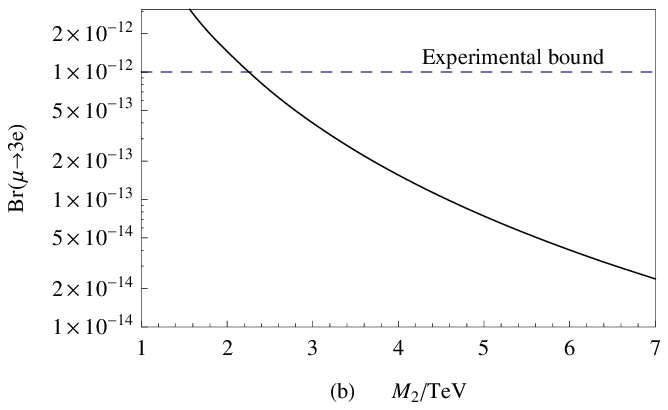}
\end{minipage}
\caption[]{Branching ratios of the LFV process vary with $M_2$ for (a) $\mu\rightarrow e\gamma$ and (b)  $\mu\rightarrow 3e$.}
\label{fig-LFV}
\end{figure}

In ref.~\cite{Zhang}, we had analyzed the LFV processes $\mu\rightarrow e\gamma$ and $\mu\rightarrow 3e$ without considering the 125 GeV Higgs. Here, constrained the 125 GeV Higgs, we also present the branching ratios of $\mu\rightarrow e\gamma$ and $\mu\rightarrow 3e$ versus $M_2$ in figure~\ref{fig-LFV}.  Similar to the case of the $\mu-e$ conversion rates in nuclei, the evaluations on the branching ratios decrease with increasing of $M_2$. As $M_2 \le 2\:{\rm{TeV}}$, the theoretical evaluations easily exceed the upper experimental bound. Differing from LFV processes which are researched in the Bilinear R-parity Violation model~\cite{BRpV-LFV}, the large VEVs of right-handed sneutrinos in the $\mu\nu$SSM induce new sources for lepton-flavor violation. So, here the branching ratios of $\mu\rightarrow e\gamma$ and $\mu\rightarrow 3e$  can easily reach upper experimental bound $5.7\times10^{-13}$~\cite{MEG} and  $1.0\times10^{-12}$~\cite{SINDRUM}, respectively. In the future, a new upgraded MEG experiment significantly improves the sensitivity reach with a goal of being able to detect the $\mu\rightarrow e\gamma$ decay at a level of $10^{-14}$~\cite{MEG2}. And the future experimental sensitivity of $\mu\rightarrow 3e$ decay will be $10^{-16}$~\cite{Mu3e}, which can detected the rare decay in the $\mu\nu$SSM.

\section{Conclusions\label{sec:6}}
Besides the superfields of the MSSM, the $\mu\nu$SSM introduces three exotic right-handed sneutrinos $\hat{\nu}_i^c$ to solve the $\mu$ problem of the MSSM. In the $\mu\nu$SSM, exotic right-handed sneutrinos which the vacuum expectation values are nonzero induce new sources for lepton-flavor violation. Additionally, we analyse the radiative correction to the SM-like Higgs in the $\mu\nu$SSM. Constrained by the ATLAS and CMS data, the numerical result indicate that the experimental data favor small $\tan \beta$.

With the 125 GeV Higgs, we analyze the $\mu-e$ conversion processes within the $\mu\nu$SSM, simultaneously considering the updated experimental data on neutrino oscillations. Numerical results indicate that the new physics corrections dominate the evaluations on the $\mu-e$ conversion rates in nuclei in some parameter space of the $\mu\nu$SSM. And we also analyse the LFV processes $\mu\rightarrow e\gamma$ and $\mu\rightarrow 3e$. The theoretical predictions on the $\mu-e$ conversion rates in nuclei $_{22}^{48}\rm{Ti}$ and $_{\:79}^{197}\rm{Au}$ and the branching ratios of $\mu\rightarrow e\gamma$ and $\mu\rightarrow 3e$ can easily reach the present experimental upper bounds and be detected in near future.

\begin{acknowledgments}
\indent\indent
The work has been supported by the National Natural Science Foundation of China (NNSFC)
with Grant No. 11275036, No. 11047002 and Natural Science Fund of Hebei University
with Grant No. 2011JQ05, No. 2012-242.
\end{acknowledgments}

\appendix

\section{Minimization of the potential\label{appendix-mini}}

First, the eight minimization conditions of the tree-level neutral scalar potential are given below:
\begin{eqnarray}
&&0=m_{{H_d}}^2 \upsilon_d + \frac{G^2}{4}( \upsilon_d^2 - \upsilon_u^2 +\upsilon_{\nu_i}\upsilon_{\nu_i}) \upsilon_d  - (A_\lambda \lambda)_i {\upsilon_u} \upsilon_{\nu_i^c} - {\lambda _j}{\kappa _{ijk}}{\upsilon_u}\upsilon_{\nu_i^c} \upsilon_{\nu_k^c}  \nonumber\\
&&\qquad  + \:({\lambda _i}{\lambda _j}\upsilon_{\nu_i^c}\upsilon_{\nu_j^c}  + {\lambda _i}{\lambda _i}\upsilon_u^2){\upsilon_d}  - {Y_{{\nu_{ij}}}}\upsilon_{\nu_i}({\lambda _k}\upsilon_{\nu_k^c}\upsilon_{\nu_j^c} + {\lambda _j}\upsilon_u^2),\\
&&0=m_{{H_u}}^2{\upsilon_u} - \frac{{G^2}}{4}(\upsilon_d^2 - \upsilon_u^2 +\upsilon_{\nu_i}\upsilon_{\nu_i})\upsilon_u  + {(A_\nu Y_\nu)}_{ij}\upsilon_{\nu_i}\upsilon_{\nu_j^c} - (A_\lambda \lambda)_i {\upsilon_d} \upsilon_{\nu_i^c}  \nonumber\\
&&\qquad  + \:({\lambda _i}{\lambda _j}\upsilon_{\nu_i^c}\upsilon_{\nu_j^c}  + {\lambda _i}{\lambda _i}\upsilon_u^2){\upsilon_u} + {Y_{{\nu_{ij}}}}\upsilon_{\nu_i}({\kappa _{ljk}}\upsilon_{\nu_l^c} \upsilon_{\nu_k^c} - 2 {\lambda _j}\upsilon_d \upsilon_u )  \nonumber\\
&&\qquad - \: {\lambda _j}{\kappa _{ijk}}{\upsilon_d}\upsilon_{\nu_i^c} \upsilon_{\nu_k^c} +  ( Y_{\nu_{ki}}Y_{\nu_{kj}}\upsilon_{\nu_i^c}\upsilon_{\nu_j^c}  + Y_{\nu_{ik}}Y_{\nu_{jk}}\upsilon_{\nu_i}\upsilon_{\nu_j} )  \upsilon_u  ,\\
&&0=m_{{{\tilde L}_{ij}}}^2\upsilon_{\nu_j} + \frac{{G^2}}{4}(\upsilon_d^2 - \upsilon_u^2 +\upsilon_{\nu_j}\upsilon_{\nu_j})\upsilon_{\nu_i}   + {(A_\nu Y_\nu)}_{ij}\upsilon_u\upsilon_{\nu_j^c} + {Y_{{\nu_{il}}}}{\kappa _{ljk}}\upsilon_u \upsilon_{\nu_j^c} \upsilon_{\nu_k^c}   \quad \nonumber\\
&&\qquad - \:  {Y_{{\nu_{ij}}}}{\lambda _k}\upsilon_{\nu_j^c} \upsilon_{\nu_k^c}  \upsilon_d -{Y_{{\nu_{ij}}}}{\lambda _j}\upsilon_u^2  \upsilon_d + {Y_{{\nu_{ij}}}}{Y_{{\nu_{lk}}}}\upsilon_{\nu_l}\upsilon_{\nu_j^c} \upsilon_{\nu_k^c} + \: {Y_{{\nu_{ik}}}}{Y_{{\nu_{jk}}}}\upsilon_u^2\upsilon_{\nu_j} ,\\
&&0=m_{\tilde \nu_{ij}^c}^2 \upsilon_{\nu_j^c} + {(A_\nu Y_\nu)}_{ji}\upsilon_{\nu_j}\upsilon_u - (A_\lambda \lambda)_i{\upsilon_d}{\upsilon_u} +{( A_\kappa \kappa)}_{ijk} \upsilon_{\nu_j^c} \upsilon_{\nu_k^c}  -  2{\lambda _j}{\kappa _{ijk}}{\upsilon_d}{\upsilon_u}\upsilon_{\nu_k^c} \nonumber\\
&&\qquad + \: {\lambda _i}{\lambda _j}\upsilon_{\nu_j^c}(\upsilon_d^2  + \upsilon_u^2) + 2{\kappa _{lim}}{\kappa _{ljk}} \upsilon_{\nu_m^c} \upsilon_{\nu_j^c} \upsilon_{\nu_k^c} +  2{Y_{{\nu_{jk}}}}{\kappa _{ikl}}{\upsilon_u}\upsilon_{\nu_j} \upsilon_{\nu_l^c} \nonumber\\
&&\qquad -\: {Y_{{\nu_{ji}}}}{\lambda _k}\upsilon_{\nu_j} \upsilon_{\nu_k^c}{\upsilon_d} - {Y_{{\nu_{kj}}}}{\lambda _i}\upsilon_{\nu_k} \upsilon_{\nu_j^c} {\upsilon_d} + {Y_{{\nu_{ji}}}}{Y_{{\nu_{lk}}}}\upsilon_{\nu_j}\upsilon_{\nu_l}\upsilon_{\nu_k^c} + {Y_{{\nu_{ki}}}}{Y_{{\nu_{kj}}}}\upsilon_u^2\upsilon_{\nu_j^c} ,
\end{eqnarray}
where $G^2=g_1^2+g_2^2$ and $g_1 c_{_W} =g_2 s_{_W}=e$.

\section{Mass Matrices\label{appendix-mass}}
In this appendix, we give the mass matrices in the $\mu\nu$SSM.

\subsection{Scalar mass matrices}
For this subsection, we use the indices $i,j,k,l,m=1,2,3$ and $\alpha=1,\ldots,8$.

\subsubsection{CP-even neutral scalars}
The quadratic potential includes
\begin{eqnarray}
{V_{quadratic}} = \frac{1}{2} {S'^T}M_S^2S' +  \cdots ,
\end{eqnarray}
where ${S'^T} = ({h_d},{h_u},{(\tilde \nu_i)^\Re},{({\tilde \nu_i^c})^\Re})$ is in the unrotated basis. And the concrete expressions for the independent coefficients of $M_S^2$ are given below:

\begin{eqnarray}
&&M_{h_d h_d}^2 = m_{H_d}^2 + \frac{G^2}{4}(3\upsilon_d^2-\upsilon_u^2+\upsilon_{\nu_i}\upsilon_{\nu_i})+\lambda_i \lambda_j \upsilon_{\nu_i^c}\upsilon_{\nu_j^c}+\lambda_i \lambda_i \upsilon_u^2 , \\
&&M_{h_u h_u}^2 = m_{H_u}^2 - \frac{G^2}{4}(\upsilon_d^2 - 3\upsilon_u^2 + \upsilon_{\nu_i}\upsilon_{\nu_i})+\lambda_i \lambda_j \upsilon_{\nu_i^c}\upsilon_{\nu_j^c}+\lambda_i \lambda_i \upsilon_d^2 \nonumber\\
&&\qquad\,\qquad - \: 2Y_{\nu_{ij}}\lambda_j \upsilon_d \upsilon_{\nu_i} + Y_{\nu_{ki}}Y_{\nu_{kj}}\upsilon_{\nu_i^c}\upsilon_{\nu_j^c}  + Y_{\nu_{ik}}Y_{\nu_{jk}}\upsilon_{\nu_i}\upsilon_{\nu_j} , \\
&&M_{h_d h_u}^2 = -(A_\lambda \lambda)_i \upsilon_{\nu_i^c} - \frac{G^2}{2}\upsilon_d \upsilon_u +2 \lambda_i \lambda_i \upsilon_d \upsilon_u  -\lambda_k \kappa_{ijk}\upsilon_{\nu_i^c}\upsilon_{\nu_j^c} \nonumber\\
&&\qquad\,\qquad - \: 2Y_{\nu_{ij}} \lambda_j \upsilon_u \upsilon_{\nu_i} , \\
&&M_{h_d (\tilde \nu_i)^\Re}^2 = \frac{G^2}{2}\upsilon_d \upsilon_{\nu_i} - Y_{\nu_{ij}}(\lambda_j \upsilon_u^2 + \lambda_k \upsilon_{\nu_k^c} \upsilon_{\nu_j^c}) ,\\
&&M_{h_u (\tilde \nu_i)^\Re}^2 = - \frac{G^2}{2}\upsilon_u \upsilon_{\nu_i} + {(A_\nu Y_\nu)}_{ij} \upsilon_{\nu_j^c} - 2 Y_{\nu_{ij}}\lambda_j \upsilon_d \upsilon_u + Y_{\nu_{ik}} \kappa_{ljk} \upsilon_{\nu_l^c}\upsilon_{\nu_j^c} \nonumber\\
&&\qquad\,\qquad\quad +\: 2 Y_{\nu_{ij}} Y_{\nu_{kj}}  \upsilon_u \upsilon_{\nu_k} , \\
&&M_{h_d (\tilde \nu_i^c)^\Re}^2 = -(A_\lambda \lambda)_i \upsilon_u +2\lambda_i \lambda_j \upsilon_d \upsilon_{\nu_j^c} - 2 \lambda_k \kappa_{ijk}\upsilon_u \upsilon_{\nu_j^c} \nonumber\\
&&\qquad\,\qquad\quad - \:( Y_{\nu_{ji}} \lambda_k  + Y_{\nu_{jk}} \lambda_i ) \upsilon_{\nu_j} \upsilon_{\nu_k^c} , \\
&&M_{h_u (\tilde \nu_i^c)^\Re}^2 = -(A_\lambda \lambda)_i \upsilon_d + {(A_\nu Y_\nu)}_{ji} \upsilon_{\nu_j} +2\lambda_i \lambda_j \upsilon_u\upsilon_{\nu_j^c} - 2 \lambda_k \kappa_{ijk}\upsilon_d \upsilon_{\nu_j^c} \qquad\;\;\; \nonumber\\
&&\qquad\,\qquad\quad + \: 2Y_{\nu_{jk}} \kappa_{ilk} \upsilon_{\nu_j}\upsilon_{\nu_l^c} +2Y_{\nu_{jk}} Y_{\nu_{ji}}  \upsilon_u \upsilon_{\nu_k^c} , \\
&&M_{(\tilde \nu_i)^\Re (\tilde \nu_j)^\Re}^2 = m_{\tilde L_{ij}}^2 +  \frac{G^2}{2}\upsilon_{\nu_i} \upsilon_{\nu_j} + \frac{G^2}{4}(\upsilon_d^2 - \upsilon_u^2 + \upsilon_{\nu_k}\upsilon_{\nu_k})\delta_{ij} \nonumber\\
&&\qquad\qquad\qquad\, +\: Y_{\nu_{ik}} Y_{\nu_{jk}}  \upsilon_u^2 + Y_{\nu_{ik}} Y_{\nu_{jl}}  \upsilon_{\nu_k^c}\upsilon_{\nu_l^c} , \\
&&M_{(\tilde \nu_i)^\Re (\tilde \nu_j^c)^\Re }^2 = {(A_\nu Y_\nu)}_{ij}\upsilon_u - (Y_{\nu_{ij}} \lambda_k + Y_{\nu_{ik}}\lambda_j ) \upsilon_d \upsilon_{\nu_k^c} + 2 Y_{\nu_{ik}} \kappa_{jlk} \upsilon_u \upsilon_{\nu_l^c}
\nonumber\\
&&\qquad\qquad\qquad\, +\:(Y_{\nu_{ij}} Y_{\nu_{kl}} +Y_{\nu_{il}} Y_{\nu_{kj}} ) \upsilon_{\nu_k} \upsilon_{\nu_l^c}  , \\
&&M_{(\tilde \nu_i^c)^\Re (\tilde \nu_j^c)^\Re }^2 =  m_{\tilde \nu_{ij}^c}^2 + 2 {(A_\kappa \kappa)}_{ijk} \upsilon_{\nu_k^c} - 2\lambda_k \kappa_{ijk} \upsilon_d \upsilon_u + \lambda_i \lambda_j ( \upsilon_d^2 + \upsilon_u^2) \nonumber\\
&&\qquad\qquad\qquad\, +\:(2\kappa_{ijk}\kappa_{lmk}+4\kappa_{ilk}\kappa_{jmk}) \upsilon_{\nu_l^c}\upsilon_{\nu_m^c} + 2Y_{\nu_{lk}} \kappa_{ijk} \upsilon_u \upsilon_{\nu_l} \nonumber\\
&&\qquad\qquad\qquad\, -\: (Y_{\nu_{kj}}\lambda_i + Y_{\nu_{ki}}\lambda_j)\upsilon_d \upsilon_{\nu_k} +Y_{\nu_{ki}}(Y_{\nu_{kj}} \upsilon_u^2 + Y_{\nu_{lj}}\upsilon_{\nu_k}\upsilon_{\nu_l}) .
\end{eqnarray}

Using an $8\times8$ unitary matrix $R_S$ to diagonalize the mass matrix $M_S^2$
\begin{eqnarray}
R_S^TM_S^2{R_S} = {(M_S^{diag})^2},
\end{eqnarray}
$S'_\alpha$ can be rotated to the mass eigenvectors $S_\alpha$:
\begin{eqnarray}
{h_d} = R_S^{1\alpha }{S_\alpha },\quad {h_u} = R_S^{2\alpha }{S_\alpha },\quad {(\tilde \nu_i)^\Re} = R_S^{(2 + i)\alpha }{S_\alpha },\quad {({\tilde \nu_i^c})^\Re} = R_S^{(5 + i)\alpha }{S_\alpha }.
\end{eqnarray}

\subsubsection{CP-odd neutral scalars}
In the unrotated basis ${P'^T} = ({P_d},{P_u},{(\tilde \nu_i)^\Im},{({\tilde \nu_i^c})^\Im})$, one can give the quadratic potential
\begin{eqnarray}
{V_{quadratic}} = \frac{1}{2}{P'^T}M_P^2P' +  \cdots ,
\end{eqnarray}
and the concrete expressions for the independent coefficients of $M_P^2$

\begin{eqnarray}
&&M_{P_d P_d}^2 = m_{H_d}^2 + \frac{G^2}{4}(\upsilon_d^2-\upsilon_u^2+\upsilon_{\nu_i}\upsilon_{\nu_i})+\lambda_i \lambda_j \upsilon_{\nu_i^c}\upsilon_{\nu_j^c}+\lambda_i \lambda_i \upsilon_u^2  , \\
&&M_{P_u P_u}^2 = m_{H_u}^2 - \frac{G^2}{4}(\upsilon_d^2 - \upsilon_u^2 + \upsilon_{\nu_i}\upsilon_{\nu_i})+\lambda_i \lambda_j \upsilon_{\nu_i^c}\upsilon_{\nu_j^c}+\lambda_i \lambda_i \upsilon_d^2 \qquad\qquad\quad\; \nonumber\\
&&\qquad\:\qquad - \: 2Y_{\nu_{ij}}\lambda_j \upsilon_d \upsilon_{\nu_i} + Y_{\nu_{ki}}Y_{\nu_{kj}}\upsilon_{\nu_i^c}\upsilon_{\nu_j^c}  + Y_{\nu_{ik}}Y_{\nu_{jk}}\upsilon_{\nu_i}\upsilon_{\nu_j} , \\
&&M_{P_d P_u}^2 = (A_\lambda \lambda)_i \upsilon_{\nu_i^c} + \lambda_k \kappa_{ijk}\upsilon_{\nu_i^c}\upsilon_{\nu_j^c}  , \\
&&M_{P_d (\tilde \nu_i)^\Im}^2 =  - Y_{\nu_{ij}}(\lambda_j \upsilon_u^2 + \lambda_k \upsilon_{\nu_k^c} \upsilon_{\nu_j^c})  ,\\
&&M_{P_u (\tilde \nu_i)^\Im}^2 = - {(A_\nu Y_\nu)}_{ij} \upsilon_{\nu_j^c} -  Y_{\nu_{ik}} \kappa_{ljk} \upsilon_{\nu_l^c}\upsilon_{\nu_j^c}  , \\
&&M_{P_d (\tilde \nu_i^c)^\Im}^2 = (A_\lambda \lambda)_i \upsilon_u  - 2 \lambda_k \kappa_{ijk}\upsilon_u \upsilon_{\nu_j^c}  - ( Y_{\nu_{ji}} \lambda_k  - Y_{\nu_{jk}} \lambda_i ) \upsilon_{\nu_j} \upsilon_{\nu_k^c} , \\
&&M_{P_u (\tilde \nu_i^c)^\Im}^2 = (A_\lambda \lambda)_i \upsilon_d - {(A_\nu Y_\nu)}_{ji} \upsilon_{\nu_j} - 2( \lambda_k \kappa_{ilk}\upsilon_d   -  Y_{\nu_{jk}} \kappa_{ilk} \upsilon_{\nu_j})\upsilon_{\nu_l^c} , \\
&&M_{(\tilde \nu_i)^\Im (\tilde \nu_j)^\Im}^2 = m_{\tilde L_{ij}}^2 + \frac{G^2}{4}(\upsilon_d^2 - \upsilon_u^2 + \upsilon_{\nu_k}\upsilon_{\nu_k})\delta_{ij}  + Y_{\nu_{ik}} Y_{\nu_{jk}}  \upsilon_u^2 \nonumber\\
&&\qquad\qquad\qquad\, + \: Y_{\nu_{ik}} Y_{\nu_{jl}}  \upsilon_{\nu_k^c}\upsilon_{\nu_l^c} , \\
&&M_{(\tilde \nu_i)^\Im (\tilde \nu_j^c)^\Im}^2 = -{(A_\nu Y_\nu)}_{ij}\upsilon_u + (Y_{\nu_{ij}} \lambda_k - Y_{\nu_{ik}}\lambda_j ) \upsilon_d \upsilon_{\nu_k^c} + 2 Y_{\nu_{il}} \kappa_{jlk} \upsilon_u \upsilon_{\nu_k^c}
\nonumber\\
&&\qquad\qquad\qquad\, -\:(Y_{\nu_{ij}} Y_{\nu_{kl}} - Y_{\nu_{il}} Y_{\nu_{kj}} ) \upsilon_{\nu_k} \upsilon_{\nu_l^c}  , \\
&&M_{(\tilde \nu_i^c)^\Im (\tilde \nu_j^c)^\Im}^2 =  m_{\tilde \nu_{ij}^c}^2 - 2 {(A_\kappa \kappa)}_{ijk} \upsilon_{\nu_k^c} + 2\lambda_k \kappa_{ijk} \upsilon_d \upsilon_u + \lambda_i \lambda_j ( \upsilon_d^2 + \upsilon_u^2) \nonumber\\
&&\qquad\qquad\qquad\, -\:(2\kappa_{ijk}\kappa_{lmk}-4\kappa_{imk}\kappa_{ljk}) \upsilon_{\nu_l^c}\upsilon_{\nu_m^c} - 2Y_{\nu_{lk}} \kappa_{ijk} \upsilon_u \upsilon_{\nu_l} \nonumber\\
&&\qquad\qquad\qquad\, -\: (Y_{\nu_{kj}}\lambda_i + Y_{\nu_{ki}}\lambda_j)\upsilon_d \upsilon_{\nu_k} +Y_{\nu_{ki}}(Y_{\nu_{kj}} \upsilon_u^2 + Y_{\nu_{lj}}\upsilon_{\nu_k}\upsilon_{\nu_l}) .
\end{eqnarray}

We can use an $8\times8$ unitary matrix $R_P$ to diagonalize the mass matrix $M_P^2$
\begin{eqnarray}
R_P^TM_P^2{R_P} = {(M_P^{diag})^2}.
\end{eqnarray}
By unitary matrix $R_P$, $P'_\alpha$ can be rotated to the mass eigenvectors $P_\alpha$:
\begin{eqnarray}
{P_d} = R_P^{1\alpha }{P_\alpha },\quad {P_u} = R_P^{2\alpha }{P_\alpha },\quad {(\tilde \nu_i)^\Im} = R_P^{(2 + i)\alpha }{P_\alpha },\quad {({\tilde \nu_i^c})^\Im} = R_P^{(5 + i)\alpha }{P_\alpha }.
\end{eqnarray}

\subsubsection{Charged scalars}
The quadratic potential includes
\begin{eqnarray}
{V_{quadratic}} = {S'^{-T}}M_{S^\pm}^2S'^+ +  \cdots ,
\end{eqnarray}
where ${S'^{ \pm T}} = (H_d^ \pm ,H_u^ \pm ,\tilde e_{L_i}^ \pm ,\tilde e_{R_i}^ \pm )$ is in the unrotated basis, $\tilde e_{L_i}^- \equiv \tilde e_i$ and $\tilde e_{R_i}^+ \equiv \tilde e_i^c$. The expressions for the independent coefficients of $M_{S^\pm}^2$ are given in detail below:

\begin{eqnarray}
&&M_{H_d^\pm H_d^\pm}^2 = m_{H_d}^2 + \frac{g_2^2}{2}(\upsilon_u^2-\upsilon_{\nu_i}\upsilon_{\nu_i}) + \frac{G^2}{4}(\upsilon_d^2-\upsilon_u^2+\upsilon_{\nu_i}\upsilon_{\nu_i})+\lambda_i \lambda_j \upsilon_{\nu_i^c}\upsilon_{\nu_j^c} \quad \nonumber\\
&&\qquad\qquad\;\;\,\; + \: Y_{e_{ik}}Y_{e_{jk}}\upsilon_{\nu_i}\upsilon_{\nu_j} , \\
&&M_{H_u^\pm H_u^\pm}^2 = m_{H_u}^2 + \frac{g_2^2}{2}(\upsilon_d^2+\upsilon_{\nu_i}\upsilon_{\nu_i}) - \frac{G^2}{4}(\upsilon_d^2 - \upsilon_u^2 + \upsilon_{\nu_i}\upsilon_{\nu_i})+\lambda_i \lambda_j \upsilon_{\nu_i^c}\upsilon_{\nu_j^c} \nonumber\\
&&\qquad\qquad\;\;\,\; + \: Y_{\nu_{ik}}Y_{\nu_{ij}}\upsilon_{\nu_j^c}\upsilon_{\nu_k^c}  , \\
&&M_{H_d^\pm H_u^\pm}^2 = (A_\lambda \lambda)_i \upsilon_{\nu_i^c} + \frac{g_2^2}{2}\upsilon_d \upsilon_u - \lambda_i \lambda_i \upsilon_d \upsilon_u  +\lambda_k \kappa_{ijk}\upsilon_{\nu_i^c}\upsilon_{\nu_j^c} \nonumber\\
&&\qquad\qquad\;\;\,\; + \: Y_{\nu_{ij}} \lambda_j \upsilon_u \upsilon_{\nu_i} , \\
&&M_{H_d^\pm \tilde e_{L_i}^\pm}^2 = \frac{g_2^2}{2}\upsilon_d \upsilon_{\nu_i} - Y_{\nu_{ij}}\lambda_k \upsilon_{\nu_k^c} \upsilon_{\nu_j^c} -  Y_{e_{ij}}Y_{e_{kj}}\upsilon_d \upsilon_{\nu_k} ,\\
&&M_{H_u^\pm \tilde e_{L_i}^\pm}^2 = \frac{g_2^2}{2}\upsilon_u \upsilon_{\nu_i} -  {(A_\nu Y_\nu)}_{ij} \upsilon_{\nu_j^c} + Y_{\nu_{ij}}\lambda_j \upsilon_d \upsilon_u - Y_{\nu_{ij}} \kappa_{ljk} \upsilon_{\nu_l^c}\upsilon_{\nu_k^c} \nonumber\\
&&\qquad\qquad\;\;\,\; -\:  Y_{\nu_{ik}} Y_{\nu_{kj}}  \upsilon_u \upsilon_{\nu_j} , \\
&&M_{H_d^\pm \tilde e_{R_i}^\pm}^2 = -(A_e Y_e)_{ji} \upsilon_{\nu_j}  -  Y_{e_{ki}}Y_{\nu_{kj}}\upsilon_u \upsilon_{\nu_j^c} , \\
&&M_{H_u^\pm \tilde e_{R_i}^\pm}^2 = - Y_{e_{ki}} (\lambda_j \upsilon_{\nu_j^c}\upsilon_{\nu_k} +Y_{\nu_{kj}}  \upsilon_d \upsilon_{\nu_j^c}) , \\
&&M_{\tilde e_{L_i}^\pm \tilde e_{L_j}^\pm}^2 = m_{\tilde L_{ij}}^2 + \frac{1}{4}(g_1^2-g_2^2)(\upsilon_d^2 - \upsilon_u^2 + \upsilon_{\nu_k}\upsilon_{\nu_k})\delta_{ij} + \frac{g_2^2}{2}\upsilon_{\nu_i} \upsilon_{\nu_j}  \nonumber\\
&&\qquad\qquad\;\;\,\; +\: Y_{\nu_{il}} Y_{\nu_{jk}}  \upsilon_{\nu_l^c}\upsilon_{\nu_k^c} +  Y_{e_{ik}} Y_{e_{jk}}  \upsilon_d^2  , \\
&&M_{\tilde e_{L_i}^\pm \tilde e_{R_j}^\pm}^2 = {(A_e Y_e)}_{ij}\upsilon_d - Y_{e_{ij}} \lambda_k  \upsilon_u \upsilon_{\nu_k^c}  , \\
&&M_{\tilde e_{R_i}^\pm \tilde e_{R_j}^\pm}^2 =  m_{\tilde e_{ij}^c}^2 - \frac{1}{2}g_1^2(\upsilon_d^2 - \upsilon_u^2 + \upsilon_{\nu_k}\upsilon_{\nu_k})\delta_{ij}  +  Y_{e_{ki}}Y_{e_{kj}} \upsilon_d^2 \nonumber\\
&&\qquad\qquad\;\;\,\; +\: Y_{e_{li}}Y_{e_{kj}}\upsilon_{\nu_k}\upsilon_{\nu_l} .
\end{eqnarray}

Through an $8\times8$ unitary matrix $R_{S^\pm}$ to diagonalize the mass matrix $M_{S^\pm}^2$
\begin{eqnarray}
R_{S^\pm}^TM_{S^\pm}^2{R_{S^\pm}} = {(M_{S^\pm}^{diag})^2},
\end{eqnarray}
we can obtain the mass eigenvectors $S^ \pm _\alpha $:
\begin{eqnarray}
H_d^ \pm  = R_{{S^ \pm }}^{1\alpha }S_\alpha ^ \pm ,\quad H_u^ \pm = R_{{S^ \pm }}^{2\alpha }S_\alpha ^ \pm ,\quad \tilde e_{L_i}^ \pm =R_{{S^ \pm }}^{(2 + i)\alpha }S_\alpha ^ \pm ,\quad \tilde e_{R_i}^ \pm = R_{{S^ \pm }}^{(5 + i)\alpha }S_\alpha ^ \pm .
\end{eqnarray}

\subsubsection{Squarks}
In the unrotated basis $\tilde{u'}^T = (\tilde{u}_{L_i},{\tilde{u}_{R_i}}^{\ast} ) \equiv (\tilde{u}_{i},{\tilde{u}_i^{c\ast}})$  and   $\tilde{d'}^T = (\tilde{d}_{L_i},{\tilde{d}_{R_i}}^{\ast} ) \equiv (\tilde{d}_{i},{\tilde{d}_i^{c\ast}})$, the quadratic potential includes
\begin{eqnarray}
{V_{quadratic}} = \frac{1}{2}{{\tilde{u'}}^{\dag}}M_{\tilde{u}}^2{\tilde{u'}} +  \frac{1}{2}{{\tilde{d'}}^{\dag}}M_{\tilde{d}}^2\tilde{d'} .
\end{eqnarray}
The concrete expressions for the independent coefficients of $M_{\tilde{u}}^2$ and $M_{\tilde{d}}^2$ are given below:
\begin{eqnarray}
&&M_{\tilde{u}_{L_i L_j}}^2 = m_{\tilde{Q}_{ij}}^2 + \frac{1}{12}(3g_2^2-g_1^2)(\upsilon_d^2-\upsilon_u^2+\upsilon_{\nu_k}\upsilon_{\nu_k}) + Y_{u_{ik}}Y_{u_{jk}}\upsilon_u^2 ,  \\
&&M_{\tilde{u}_{R_i R_j}}^2 = m_{\tilde{u}_{ij}^c}^2 + \frac{1}{3}g_1^2(\upsilon_d^2-\upsilon_u^2+\upsilon_{\nu_k}\upsilon_{\nu_k}) + Y_{u_{ki}}Y_{u_{kj}}\upsilon_u^2 ,  \\
&&M_{\tilde{u}_{L_i R_j}}^2 = (A_u Y_u)_{ij} \upsilon_u - Y_{u_{ij}}\lambda_k \upsilon_d \upsilon_{\nu_k^c} + Y_{\nu_{lk}}Y_{u_{ij}}\upsilon_{\nu_l}\upsilon_{\nu_k^c} ,  \\
&&M_{\tilde{u}_{R_i L_j}}^2 = M_{\tilde{u}_{L_j R_i}}^2 ,
\end{eqnarray}
and
\begin{eqnarray}
&&M_{\tilde{d}_{L_i L_j}}^2 = m_{\tilde{Q}_{ij}}^2 - \frac{1}{12}(3g_2^2+g_1^2)(\upsilon_d^2-\upsilon_u^2+\upsilon_{\nu_k}\upsilon_{\nu_k}) + Y_{d_{ik}}Y_{d_{jk}}\upsilon_d^2 , \\
&&M_{\tilde{d}_{R_i R_j}}^2 = m_{\tilde{d}_{ij}^c}^2 - \frac{1}{6}g_1^2(\upsilon_d^2-\upsilon_u^2+\upsilon_{\nu_k}\upsilon_{\nu_k}) + Y_{d_{ki}}Y_{d_{kj}}\upsilon_d^2 ,  \\
&&M_{\tilde{d}_{L_i R_j}}^2 = (A_d Y_d)_{ij} \upsilon_d - Y_{d_{ij}}\lambda_k \upsilon_u \upsilon_{\nu_k^c} ,  \\
&&M_{\tilde{d}_{R_i L_j}}^2 = M_{\tilde{d}_{L_j R_i}}^2 .
\end{eqnarray}

With the diagonal mass matrix
\begin{eqnarray}
R_{q}^\dag M_{\tilde{q}}^2{R_{q}} = {(M_{\tilde{q}}^{diag})^2},\quad (q=u,d)
\end{eqnarray}
$\tilde{u'}$ and $\tilde{d'}$ can be rotated to the mass eigenvectors $U^ \pm _I$ and $D^ \pm _I$ $(I=1, \ldots, 6)$:
\begin{eqnarray}
\left\{ {\begin{array}{*{20}{c}}
   {\tilde{u}_i = R_u^{iI }U^+_I ,\quad\: \tilde{u}_i^c = R_u^{(3+i)I\ast }U^-_I ;}  \\
   {\tilde{d}_i = R_d^{iI }D^-_I ,\quad\: \tilde{d}_i^c = R_d^{(3+i)I\ast }D^+_I .}  \\
\end{array}} \right.
\end{eqnarray}

\subsection{Neutral fermion mass matrix}
Neutralinos mix with the neutrinos and in a basis ${\chi '^{ \circ T}} = \left( {{{\tilde B}^ \circ },{{\tilde W}^ \circ },{{\tilde H}_d}{\rm{,}}{{\tilde H}_u},{\nu_{R_i}},{\nu_{L{_i}}}} \right)$, one obtains the neutral fermion mass terms in the Lagrangian:
\begin{eqnarray}
 - \frac{1}{2}{\chi '^{ \circ T}}{M_n}{\chi '^ \circ } + {\rm{H.c.}}  ,
\end{eqnarray}
where
\begin{eqnarray}
{M_n} = \left( {\begin{array}{*{20}{c}}
   M & {{m^T}}  \\
   m & {{0_{3 \times 3}}}  \\
\end{array}} \right),
\end{eqnarray}
with
\begin{eqnarray}
m = \left( {\begin{array}{*{20}{c}}
   {  -\frac{g_1}{\sqrt 2 }\upsilon_{{\nu _1}}} & { \frac{g_2}{\sqrt 2 }\upsilon_{{\nu _1}}} & 0 & {{Y_{{\nu _{1i}}}}{\upsilon_{\nu _i^c}}} & {{Y_{{\nu _{11}}}}{\upsilon_u}} & {{Y_{{\nu _{12}}}}{\upsilon_u}} & {{Y_{{\nu _{13}}}}{\upsilon_u}}  \\
   {  -\frac{g_1}{\sqrt 2 }\upsilon_{{\nu _2}}} & { \frac{g_2}{\sqrt 2 }\upsilon_{{\nu _2}}} & 0 & {{Y_{{\nu _{2i}}}}{\upsilon_{\nu _i^c}}} & {{Y_{{\nu _{21}}}}{\upsilon_u}} & {{Y_{{\nu _{22}}}}{\upsilon_u}} & {{Y_{{\nu _{23}}}}{\upsilon_u}}  \\
   {  -\frac{g_1}{\sqrt 2 }\upsilon_{{\nu _3}}} & { \frac{g_2}{\sqrt 2 }\upsilon_{{\nu _3}}} & 0 & {{Y_{{\nu _{3i}}}}{\upsilon_{\nu _i^c}}} & {{Y_{{\nu _{31}}}}{\upsilon_u}} & {{Y_{{\nu _{32}}}}{\upsilon_u}} & {{Y_{{\nu _{33}}}}{\upsilon_u}}  \\
\end{array}} \right)
\end{eqnarray}
and
\begin{eqnarray}
&&M = \left( {\begin{array}{*{20}{c}}
   {{M_1}} & 0 & {\frac{-g_1}{{\sqrt 2 }}{\upsilon _d}} & {\frac{g_1}{{\sqrt 2 }}{\upsilon _u}} & 0 & 0 & 0  \\
   0 & {{M_2}} & {\frac{g_2}{{\sqrt 2 }}{\upsilon _d}} & {\frac{-g_2}{{\sqrt 2 }}{\upsilon _u}} & 0 & 0 & 0  \\
   {\frac{-g_1}{{\sqrt 2 }}{\upsilon _d}} & {\frac{g_2}{{\sqrt 2 }}{\upsilon _d}} & 0 & {-{\lambda _i}{\upsilon _{\nu _i^c}}} & { - {\lambda _1}{\upsilon _u}} & { - {\lambda _2}{\upsilon _u}} & { - {\lambda _3}{\upsilon _u}}  \\
   {\frac{g_1}{{\sqrt 2 }}{\upsilon _u}} & {\frac{-g_2}{{\sqrt 2 }}{\upsilon _u}} & {-{\lambda _i}{\upsilon _{\nu _i^c}}} & 0 & {y_1} & {y_2} & { y_3}  \\
   0 & 0 & { - {\lambda _1}{\upsilon _u}} & { y_1} & {2{\kappa _{11j}}{\upsilon _{\nu _j^c}}} & {2{\kappa _{12j}}{\upsilon _{\nu _j^c}}} & {2{\kappa _{13j}}{\upsilon _{\nu _j^c}}}  \\
   0 & 0 & { - {\lambda _2}{\upsilon _u}} & { y_2} & {2{\kappa _{21j}}{\upsilon _{\nu _j^c}}} & {2{\kappa _{22j}}{\upsilon _{\nu _j^c}}} & {2{\kappa _{23j}}{\upsilon _{\nu _j^c}}}  \\
   0 & 0 & { - {\lambda _3}{\upsilon _u}} & { y_3} & {2{\kappa _{31j}}{\upsilon _{\nu _j^c}}} & {2{\kappa _{32j}}{\upsilon _{\nu _j^c}}} & {2{\kappa _{33j}}{\upsilon _{\nu _j^c}}}  \\
\end{array}} \right)
\end{eqnarray}
where $y_i=- {\lambda _i}{\upsilon _d}+ {{Y_{{\nu _{ji}}}}{\upsilon _{{\nu _j}}} }$. Here, the submatrix $m$ is neutralino-neutrino mixing, and the submatrix $M$ is neutralino mass matrix. This $10\times10$ symmetric matrix $M_n$ can be diagonalized by a $10\times10$ unitary matrix $Z_n$:
\begin{eqnarray}
Z_n^T{M_n}{Z_n} = {M_{nd}},
\end{eqnarray}
where $M_{nd}$ is the diagonal neutral fermion mass matrix. Then, one can obtain the neutral fermion mass eigenstates:
\begin{eqnarray}
\chi _\alpha ^ \circ  = \left( {\begin{array}{*{20}{c}}
   {\kappa _\alpha ^ \circ }  \\
   { \overline{\kappa _\alpha ^ \circ} }  \\
\end{array}} \right), \quad {\alpha  = 1, \ldots, 10}
\end{eqnarray}
with
\begin{eqnarray}
\left\{ {\begin{array}{*{20}{c}}
   {{\tilde B^ \circ } = Z_n^{1\alpha }\kappa _\alpha ^ \circ ,\quad\: {\tilde H_d} = Z_n^{3\alpha }\kappa _\alpha ^ \circ ,\quad{\nu_{R_i}} = Z_n^{\left( {4 + i} \right)\alpha }\kappa _\alpha ^ \circ ,}  \\
   {{\tilde W^ \circ } = Z_n^{2\alpha }\kappa _\alpha ^ \circ ,\quad{\tilde H_u} = Z_n^{4\alpha }\kappa _\alpha ^ \circ , \quad{\nu_{L_i}} = Z_n^{\left( {7 + i} \right)\alpha }\kappa _\alpha ^ \circ .}  \\
\end{array}} \right.
\end{eqnarray}

\subsection{Charged fermion mass matrix}
Charginos mix with the charged leptons and in a basis where ${\Psi ^{ - T}} = \left( { - i{{\tilde \lambda }^ - },\tilde H_d^ - ,e_{L{_i}}^ - } \right)$ and ${\Psi ^{ + T}} = \left( { - i{{\tilde \lambda }^ + },\tilde H_u^ + ,e_{R{_i}}^+} \right)$, one can obtain the charged fermion mass terms in the Lagrangian:
\begin{eqnarray}
- {\Psi ^{ - T}}{M_c}{\Psi^+} + {\rm{H.c.}},
\end{eqnarray}
where
\begin{eqnarray}
{M_c} = \left( {\begin{array}{*{20}{c}}
   {{M_ \pm }} & b  \\
   c & {{m_l}}  \\
\end{array}} \right).
\end{eqnarray}
Here, the submatrix $M_ \pm $ is chargino mass matrix
\begin{eqnarray}
{M_ \pm } = \left( {\begin{array}{*{20}{c}}
   {{M_2}} & {g_2 {\upsilon_u}}  \\
   {g_2 {\upsilon_d}} & {{\lambda _i} \upsilon_{\nu_i^c}}  \\
\end{array}} \right).
\end{eqnarray}
And the submatrices $b$ and $c$ give rise to chargino-charged lepton mixing. They are defined as
\begin{eqnarray}
b = \left( {\begin{array}{*{20}{c}}
   0 & 0 & 0  \\
   { - {Y_{e_{i1}}} \upsilon_{\nu _i}} & { - {Y_{e_{i2}}} \upsilon_{\nu _i}} & { - {Y_{e_{i3}}} \upsilon_{\nu _i}}  \\
\end{array}} \right),
\end{eqnarray}
\begin{eqnarray}
c = \left( {\begin{array}{*{20}{c}}
   {g_2 \upsilon_{\nu _1}} & { - {Y_{\nu_{1i}}}\upsilon_{\nu_i^c}}  \\
   {g_2 \upsilon_{\nu _2}} & { - {Y_{\nu_{2i}}}\upsilon_{\nu_i^c}}  \\
   {g_2 \upsilon_{\nu _3}} & { - {Y_{\nu_{3i}}}\upsilon_{\nu_i^c}}  \\
\end{array}} \right).
\end{eqnarray}
And the submatrix $m_l$ is the charged lepton mass matrix
\begin{eqnarray}
{m_l} = \left( {\begin{array}{*{20}{c}}
   {{Y_{e_{11}}}{\upsilon_d}} & {{Y_{e_{12}}}{\upsilon_d}} & {{Y_{e_{13}}}{\upsilon_d}}  \\
   {{Y_{e_{21}}}{\upsilon_d}} & {{Y_{e_{22}}}{\upsilon_d}} & {{Y_{e_{23}}}{\upsilon_d}}  \\
   {{Y_{e_{31}}}{\upsilon_d}} & {{Y_{e_{32}}}{\upsilon_d}} & {{Y_{e_{33}}}{\upsilon_d}}  \\
\end{array}} \right).
\end{eqnarray}
This $5\times5$ mass matrix $M_c$ can be diagonalized by the $5\times5$ unitary matrices $Z_-$ and $Z_+$:
\begin{eqnarray}
Z_ - ^T{M_c}{Z_ + } = {M_{cd}},
\end{eqnarray}
where $M_{cd}$ is the diagonal charged fermion mass matrix. Then, we can obtain the charged fermion mass eigenstates:
\begin{eqnarray}
{\chi _\alpha } = \left( {\begin{array}{*{20}{c}}
   {\kappa _\alpha ^ - }  \\
   {\overline{{\kappa _\alpha ^+}}}  \\
\end{array}} \right),\quad {\alpha  = 1, \ldots, 5}
\end{eqnarray}
with
\begin{eqnarray}
\left\{ {\begin{array}{*{20}{c}}
   {{{\tilde \lambda }^ - } = iZ_ - ^{1\alpha }\kappa _\alpha ^ - ,\quad \tilde H_d^ -  = Z_ - ^{2\alpha }\kappa _\alpha ^ - ,\quad {e_{L_i}} = Z_ - ^{\left( {2 + i} \right)\alpha }\kappa _\alpha ^ - ;}  \\
   \;{{{\tilde \lambda }^ + } = iZ_ + ^{1\alpha }\kappa _\alpha ^ + ,\quad {{\tilde H}_u^+} = Z_ + ^{2\alpha }\kappa _\alpha ^ + ,\quad {e_{R_i}} = Z_+^{\left( {2 + i} \right)\alpha }\kappa _\alpha ^+.}  \\
\end{array}} \right.
\end{eqnarray}

\section{Approximate diagonalization of mass matrices \label{appendix-approximate}}

\subsection{Neutral fermion mass matrix}
If the R-parity breaking parameters are small in the sense that for~\cite{R11,Valle2}
\begin{eqnarray}
\xi  = m.{M^{ - 1}},
\end{eqnarray}
all ${\xi _{ij}} \ll 1$, one can find an approximate diagonalization of neutral fermion mass matrix. In leading order in  $\xi$, the rotation matrix $Z_n$ is given by
\begin{eqnarray}
{Z_n} = \left( {\begin{array}{*{20}{c}}
   {1 - \frac{1}{2}{\xi ^T}\xi } & { - {\xi ^T}}  \\
   \xi  & {1 - \frac{1}{2}\xi {\xi ^T}}  \\
\end{array}} \right)\left( {\begin{array}{*{20}{c}}
   V & 0  \\
   0 & {{U_\nu }}  \\
\end{array}} \right).
\label{Zn}
\end{eqnarray}
The first matrix in (\ref{Zn}) above approximately block-diagonalizes the matrix $M_n$ to the form $diag\left( {M,{m_{eff}}} \right)$, where
\begin{eqnarray}
{m_{eff}} =  -  m.{M^{ - 1}} .{m^T}.
\end{eqnarray}
The submatrices $V$ and $U_{\nu}$ respectively diagonalize $M$ and ${m_{eff}}$ in the following way:
\begin{eqnarray}
\left\{ \begin{array}{l}
 {V^T}MV = {M_d}, \\
 U_\nu ^T{m_{eff}}{U_\nu} = {m_{\nu d}}, \\
 \end{array} \right.
\end{eqnarray}
where $M_d$ and ${m_{\nu d}}$ are respectively diagonal neutralino and neutrino mass matrix.

\subsection{Charged fermion mass matrix}
Similarly to the case of the neutral fermion mass matrix discussed above, it's also possible to find an approximate diagonalization of the charged fermion mass matrix for the small R-parity breaking parameters~\cite{Valle3}. Define then,
\begin{eqnarray}
\left\{ \begin{array}{l}
 {\xi _L} = c.M_ \pm ^{ - 1} + {m_l}.{b^T}.{(M_ \pm ^{ - 1})^T}.M_ \pm ^{ - 1}; \\
 {\xi _R} = {b^T}.{(M_ \pm ^{ - 1})^T} + {m_l}^T.c.M_ \pm ^{ - 1}.{(M_ \pm ^{ - 1})^T}. \\
 \end{array} \right.
\end{eqnarray}
All $\xi_{L_{ij}} \ll 1$ and $\xi_{R_{ij}} \ll 1$, so in leading order in $\xi_L$ and $\xi_R$, the rotation matrices $Z_-$ and $Z_+$ are respectively given as
\begin{eqnarray}
{Z_-} = \left( {\begin{array}{*{20}{c}}
   {1 - \frac{1}{2}{\xi_L ^T}\xi_L } & { - {\xi_L ^T}}  \\
   \xi_L  & {1 - \frac{1}{2}\xi_L {\xi_L ^T}}  \\
\end{array}} \right)\left( {\begin{array}{*{20}{c}}
   U_- & 0  \\
   0 & {{V_- }}  \\
\end{array}} \right)
\end{eqnarray}
and
\begin{eqnarray}
{Z_+} = \left( {\begin{array}{*{20}{c}}
   {1 - \frac{1}{2}{\xi_R ^T}\xi_R } & { - {\xi_R ^T}}  \\
   \xi_R  & {1 - \frac{1}{2}\xi_R {\xi_R ^T}}  \\
\end{array}} \right)\left( {\begin{array}{*{20}{c}}
   U_+ & 0  \\
   0 & {{V_+ }}  \\
\end{array}} \right).
\end{eqnarray}
Then the matrix $M_c$ can approximately be block-diagonalized to the form $diag\left( {{M_ \pm },{m_l}} \right)$. And the submatrices $U_-,U_+$ and $V_-,V_+$ respectively diagonalize $M_\pm$ and $m_l$ in the following way:
\begin{eqnarray}
\left\{ \begin{array}{l}
 U_ - ^T{M_ \pm }{U_ + } = {M_{ \pm d}}, \\
 V_ - ^T{m_l}{V_ + } = {m_{ld}}, \\
 \end{array} \right.
\end{eqnarray}
where ${M_{ \pm d}}$ and ${m_{ld}}$ are respectively diagonal chargino and charged lepton mass matrix.

\section{Interaction Lagrangian\label{appendix-interaction}}

In this part, we give the interaction Lagrangian of the relative vertices for the LFV processes in the $\mu\nu$SSM. And we use the indices $i,j=1,\ldots,3$, $\beta,\zeta=1,\ldots,5$, $I=1,\ldots,6$, $\alpha,\rho=1,\ldots,8$ and $\eta=1,\ldots,10$.

\subsection{Charged fermion-neutral fermion-gauge boson }
One can give the interaction Lagrangian of charged fermion, neutral fermion and gauge boson
\begin{eqnarray}
&&\mathcal{L}_{int} = e F_\mu \bar{\chi}_\beta \gamma^\mu \chi_\beta + Z_\mu \bar{\chi}_\beta (C_L^{Z \chi_\zeta \bar{\chi}_\beta}\gamma^\mu P_L +  C_R^{Z \chi_\zeta \bar{\chi}_\beta}\gamma^\mu P_R) \chi_\zeta\nonumber\\
&&\qquad\quad  +\; W_\mu^+ \bar{\chi}_\eta^0 (C_L^{W \chi_\beta \bar{\chi}_\eta^0}\gamma^\mu P_L  +  C_R^{W \chi_\beta \bar{\chi}_\eta^0}\gamma^\mu P_R) \chi_\beta \nonumber\\
&&\qquad\quad  +\; W_\mu^- \bar{\chi}_\beta (C_L^{W \chi_\eta^0 \bar{\chi}_\beta}\gamma^\mu P_L +  C_R^{W \chi_\eta^0 \bar{\chi}_\beta}\gamma^\mu P_R) \chi_\eta^0 + \cdots,
\end{eqnarray}
where the coefficients are
\begin{eqnarray}
&&C_L^{Z{\chi _\zeta }{{\bar \chi }_{^\beta }}} = \frac{e}{{2{s_{_W}}{c_{_W}}}}\Big[ {( {1 - 2s_{_W}^2} ){\delta ^{\zeta \beta }} + Z{{_ - ^{1\zeta }}^ * }Z_ - ^{1\beta }} \Big], \\
&&C_R^{Z{\chi _\zeta }{{\bar \chi }_{^\beta }}} = \frac{e}{{2{s_{_W}}{c_{_W}}}}\Big[ {2Z{{_ + ^{1\zeta }}^ * }Z_ + ^{1\beta } + Z{{_ + ^{2\zeta }}^ * }Z_ + ^{2\beta } - 2s_W^2{\delta ^{\zeta \beta }}} \Big], \\
&&C_L^{W{\chi _{^\beta }}\bar \chi _\eta ^ \circ } =  - \frac{e}{{\sqrt 2 {s_{_W}}}}\Big[ \sqrt 2 Z_ - ^{1\beta }Z{{_n^{2\eta }}^ * } + Z_ - ^{2\beta }Z{{_n^{3\eta }}^ * } + Z_ - ^{(2 + i)\beta }Z{{_n^{(7 + i)\eta }}^ * } \Big], \\
&&C_R^{W{\chi _{^\beta }}\bar \chi _\eta ^ \circ } =  - \frac{e}{{\sqrt 2 {s_{_W}}}}\Big[ \sqrt 2 Z{{_ + ^{1\beta }}^ * }Z_n^{2\eta } - Z{{_ + ^{2\beta }}^ * }Z_n^{4\eta } \Big],\\
&&C_L^{W\chi _\eta ^ \circ {{\bar \chi }_{^\beta }}} = \Big[ {C_L^{W{\chi _{^\beta }}\bar \chi _\eta ^ \circ }} \Big]^ * ,\qquad  C_R^{W\chi _\eta ^ \circ {{\bar \chi }_{^\beta }}} = \Big[ {C_R^{W{\chi _{^\beta }}\bar \chi _\eta ^ \circ }} \Big]^ * .
\end{eqnarray}

\subsection{Charged scalar-gauge boson}
The interaction Lagrangian of charged scalar and gauge boson is written by
\begin{eqnarray}
\mathcal{L}_{int} = i e F_\mu S_\alpha^{-\ast}{\mathord{\buildrel{\lower3pt\hbox{$\scriptscriptstyle\leftrightarrow$}}
\over {\partial^\mu} } } S_\alpha^- + i e C^{Z S_\alpha^- S_\rho^{-\ast}} Z_\mu S_\rho^{-\ast}{\mathord{\buildrel{\lower3pt\hbox{$\scriptscriptstyle\leftrightarrow$}}
\over {\partial^\mu} } } S_\alpha^-  + \cdots,
\end{eqnarray}
where the coefficient is
\begin{eqnarray}
{C^{ZS_\alpha ^ -  S_\rho ^{ -  * }}} = \frac{e}{{2{s_{_W}}{c_{_W}}}}\Big[ ( {1 - 2s_{_W}^2} ){\delta ^{\alpha \rho }} - R{{_{{S^ \pm }}^{(5 + i)\alpha }}^ * }R_{{S^ \pm }}^{(5 + i)\rho } \Big].
\end{eqnarray}

\subsection{Charged fermion-neutral fermion-scalar}
The interaction Lagrangian of charged fermion, neutral fermion and scalar can be similarly written as
\begin{eqnarray}
&&\mathcal{L}_{int} = S_\alpha \bar{\chi}_\zeta (C_L^{{S_\alpha }{\chi _\beta }{{\bar \chi }_\zeta }}{P_L} + C_R^{{S_\alpha }{\chi _\beta }{{\bar \chi }_\zeta }}{P_R}) \chi_\beta + P_\alpha \bar{\chi}_\zeta (C_L^{{P_\alpha }{\chi _\beta }{{\bar \chi }_\zeta }}{P_L}  \nonumber\\
&& \qquad\quad + C_R^{{P_\alpha }{\chi _\beta }{{\bar \chi }_\zeta }}P_R ) \chi_\beta + S_\alpha^- \bar{\chi}_\beta (C_L^{S_\alpha ^ - \chi _\eta ^ \circ {{\bar \chi }_\beta }}{P_L} + C_R^{S_\alpha ^ - \chi _\eta ^ \circ {{\bar \chi }_\beta }}{P_R} ) \chi_\eta^0  \nonumber\\
&& \qquad\quad + S_\alpha^{-\ast} \bar{\chi}_\eta^0 (C_L^{S_\alpha ^{-\ast} {\chi _\beta }\bar \chi _\eta ^ \circ }{P_L} + C_R^{S_\alpha ^{-\ast} {\chi _\beta }\bar \chi _\eta ^ \circ }{P_R} ) \chi_\beta  + \cdots,
\end{eqnarray}
where the coefficients are
\begin{eqnarray}
&&C_L^{{S_\alpha }{\chi _\beta }{{\bar \chi }_\zeta }} =   \frac{-e}{{{\sqrt{2}s_{_W}}}}\Big[ R_S^{2\alpha }Z_ - ^{1\beta }Z_ + ^{2\zeta } + R_S^{1\alpha }Z_ - ^{2\beta }Z_ + ^{1\zeta } + R_S^{(5 + i)\alpha }Z_ - ^{(2 + i)\beta }Z_ + ^{1\zeta } \Big]  \nonumber\\
&&\qquad\qquad\quad + \,\frac{1}{\sqrt{2}} {Y_{e_{ij}}}\Big[ R_S^{(5 + i)\alpha }Z_ - ^{1\beta }Z_ + ^{(2 + j)\zeta } - R_S^{1\alpha }Z_ - ^{(2 + i)\beta }Z_ + ^{(2 + j)\zeta }  \Big] \nonumber\\
&&\qquad\qquad\quad - \,\frac{1}{\sqrt{2}}{Y_{\nu_{ij}}}R_S^{(2 + j)\alpha }Z_ - ^{(2 + i)\beta }Z_ + ^{2\zeta } - \frac{1}{\sqrt{2}}{\lambda _i}R_S^{(2 + i)\alpha }Z_ - ^{2\beta }Z_ + ^{2\zeta }  ,
\end{eqnarray}
\begin{eqnarray}
&&C_L^{{P_\alpha }{\chi _\beta }{{\bar \chi }_\zeta }} = \frac{{ie}}{{{\sqrt{2}s_{_W}}}}\Big[R_P^{2\alpha }Z_ - ^{1\beta }Z_ + ^{2\zeta } + R_P^{1\alpha }Z_ - ^{2\beta }Z_ + ^{1\zeta } + R_P^{(5 + i)\alpha }Z_ - ^{(2 + i)\beta }Z_ + ^{1\zeta }\Big]  \nonumber\\
&&\qquad\qquad\quad  + \, \frac{i}{\sqrt{2}}{Y_{{e_{ij}}}}\Big[ R_P^{(5 + i)\alpha }Z_ - ^{1\beta }Z_ + ^{(2 + j)\zeta } - R_P^{1\alpha }Z_ - ^{(2 + i)\beta }Z_ + ^{(2 + j)\zeta } \Big] \nonumber\\
&&\qquad\qquad\quad  - \, \frac{i}{\sqrt{2}}{Y_{{\nu _{ij}}}}R_P^{(2 + j)\alpha }Z_ - ^{(2 + i)\beta }Z_ + ^{2\zeta } - \frac{i}{\sqrt{2}}{\lambda _i}R_P^{(2 + i)\alpha }Z_ - ^{2\beta }Z_ + ^{2\zeta } , \\
&&C_L^{S_\alpha^- \chi _\eta^0 {{\bar{\chi}}_\beta }} =   \frac{-e}{{\sqrt{2} {s_W}{c_W}}}R{_{{S^\pm }}^{2\alpha \ast } }Z_+^{2\beta} \Big[ {{c_W}Z_n^{2\eta } + {s_W}Z_n^{1\eta }} \Big]  - \frac{e}{{{s_W}}}R{_{{S^ \pm }}^{2\alpha\ast } }Z_ + ^{1\beta }Z_n^{4\eta }  \nonumber\\
&&\qquad\qquad\quad - \frac{{\sqrt{2} e}}{{{s_W}}}R{_{{S^\pm }}^{(5 + i)\alpha\ast } }Z_ + ^{(2 + i)\beta }Z_n^{1\eta } + {Y_{\nu_{ij}}}R_{{S^ \pm }}^{(2 + i)\alpha }Z_ + ^{2\beta }Z_n^{(4 + j)\eta } \nonumber\\
&&\qquad\qquad\quad  + \, {Y_{e_{ij}}}Z_ + ^{(2 + j)\beta } \Big[ R_{{S^ \pm }}^{1\alpha }Z_n^{(7 + i)\eta } - R_{{S^ \pm }}^{(2 + i)\alpha }Z_n^{3\eta } \Big] - {\lambda _i}R_{{S^ \pm }}^{1\alpha }Z_ + ^{2\beta }Z_n^{(4 + i)\eta },\\
&&C_L^{S_\alpha ^{-\ast} {\chi _\beta }\bar \chi _\eta ^ \circ } =   \frac{e}{{\sqrt 2 {s_W}{c_W}}}\Big[ R{{_{{S^ \pm }}^{1\alpha\ast }} }Z_ - ^{2\beta } + R{{_{{S^ \pm }}^{(2 + i)\alpha }}^ * }Z_ - ^{(2 + i)\beta }\Big]\Big[ {c_W}Z_n^{2\eta } + {s_W}Z_n^{1\eta }\Big] \nonumber\\
&&\qquad\qquad\quad\;\; - \frac{e}{{{s_W}}}Z_ - ^{1\beta }\Big[ R{{_{{S^ \pm }}^{1\alpha\ast }} }Z_n^{3\eta } + R{{_{{S^ \pm }}^{(2 + i)\alpha\ast }} }Z_n^{(7 + i)\eta }\Big] + {Y_{\nu_{ij}}}R_{{S^ \pm }}^{2\alpha }Z_ - ^{(2 + i)\beta }Z_n^{(4 + j)\eta }\nonumber\\
&&\qquad\qquad\quad\;\; +\: {Y_{{e_{ij}}}}R_{{S^ \pm }}^{(5 + j)\alpha }\Big[ Z_ - ^{2\beta }Z_n^{(7 + i)\eta } - Z_ - ^{(2 + i)\beta }Z_n^{3\eta } \Big] - {\lambda _i}R_{{S^ \pm }}^{2\alpha }Z_ - ^{2\beta }Z_n^{(4 + i)\eta },\\
&&C_R^{{S_\alpha }{\chi _\beta }{{\bar \chi }_\zeta }} = \Big[ {C_L^{{S_\alpha }{\chi _\zeta }{{\bar \chi }_\beta }}} \Big]^ * ,\qquad\quad    C_R^{{P_\alpha }{\chi _\beta }{{\bar \chi }_\zeta }} = \Big[ {C_L^{{P_\alpha }{\chi _\zeta }{{\bar \chi }_\beta }}} \Big]^ * ,\\
&&C_R^{S_\alpha ^ - \chi _\eta ^ \circ {{\bar \chi }_\beta }} = \Big[ {C_L^{S_\alpha ^{-\ast} {\chi _\beta }\bar \chi _\eta ^0 }} \Big]^ * , \qquad  C_R^{S_\alpha ^{-\ast} {\chi _\beta }\bar \chi _\eta ^ \circ } =\Big[ {C_L^{S_\alpha ^ - \chi _\eta ^ \circ {{\bar \chi }_\beta }}} \Big]^ * .
\end{eqnarray}

\subsection{Quark-squark-fermion}
The interaction Lagrangian of quark, squark and fermion is similarly written by
\begin{eqnarray}
&&\mathcal{L}_{int} = \Big[ U_I^+ \bar{u}_i (C_L^{U_I^+ {\chi _\alpha^0}\bar{u}_i}{P_L} + C_R^{U_I^+ {\chi _\alpha^0}\bar{u}_i}{P_R}) \chi _\alpha^0 + D_I^- \bar{d}_i (C_L^{D_I^- {\chi _\alpha^0}\bar{d}_i}{P_L} \nonumber\\
&& \qquad\quad\;\; + \: C_R^{D_I^- {\chi _\alpha^0}\bar{d}_i}{P_R}) \chi _\alpha^0    + U_I^+ \bar{d}_i (C_L^{U_I^+ {\chi _\alpha}\bar{d}_i}{P_L} + C_R^{U_I^+ {\chi _\alpha}\bar{d}_i}{P_R}) \chi _\alpha  \nonumber\\
&& \qquad\quad\;\; + \: D_I^- \bar{\chi}_\alpha (C_L^{D_I^- u_i^c \bar{\chi}_\alpha}{P_L} + C_R^{D_I^- u_i^c \bar{\chi}_\alpha}{P_R})u_i^c  \Big] + {\rm{H.c.}}.
\end{eqnarray}
And the coefficients are
\begin{eqnarray}
&&C_L^{U_I^+ {\chi _\alpha^0}\bar{u}_i} =  \frac{2\sqrt{2}e}{3{c_{_W}}} Z_n^{1\alpha}R_u^{(3+i)I } - {Y_{u_{ji}}} Z_n^{4\alpha}R_u^{jI }, \\
&&C_R^{U_I^+ {\chi _\alpha^0}\bar{u}_i} = \frac{-e}{\sqrt{2}{s_{_W}}{c_{_W}}} (\frac{1}{3}Z_n^{1\alpha\ast}{s_{_W}}+Z_n^{2\alpha\ast}{c_{_W}})R_u^{iI } - {Y_{u_{ij}}} Z_n^{4\alpha\ast}R_u^{(3+j)I }, \\
&&C_L^{D_I^- {\chi _\alpha^0}\bar{d}_i} =  \frac{-\sqrt{2}e}{3{c_{_W}}} Z_n^{1\alpha}R_d^{(3+i)I } - {Y_{d_{ji}}} Z_n^{3\alpha}R_d^{jI }, \\
&&C_R^{D_I^- {\chi _\alpha^0}\bar{d}_i} = \frac{-e}{\sqrt{2}{s_{_W}}{c_{_W}}} (\frac{1}{3}Z_n^{1\alpha\ast}{s_{_W}}-Z_n^{2\alpha\ast}{c_{_W}})R_d^{iI } - {Y_{d_{ij}}} Z_n^{3\alpha\ast}R_d^{(3+j)I }, \\
&&C_L^{U_I^+ {\chi _\alpha}\bar{d}_i} =  {Y_{d_{ji}}} Z_-^{2\alpha}R_u^{jI }, \\
&&C_R^{U_I^+ {\chi _\alpha}\bar{d}_i} = \frac{-e}{s_{_W}} Z_+^{1\alpha\ast}R_u^{iI } + {Y_{u_{ij}}} Z_+^{2\alpha\ast}R_u^{(3+j)I }, \\
&&C_L^{D_I^- u_i^c \bar{\chi}_\alpha} =  {Y_{u_{ji}}} Z_+^{2\alpha}R_d^{jI }, \\
&&C_R^{D_I^- u_i^c \bar{\chi}_\alpha} = \frac{-e}{s_{_W}} Z_-^{1\alpha\ast}R_d^{iI } + {Y_{d_{ij}}} Z_-^{2\alpha\ast}R_d^{(3+j)I }.
\end{eqnarray}

\section{Form factors \label{appendix-integral}}
Defining ${x_i} = \frac{{m_i^2}}{{m_W^2}}$, we can find the form factors:
\begin{eqnarray}
&&{I_1}(\textit{x}_1 , x_2 ) = \frac{1}{{16{\pi ^2}}}\Big[ \frac{{1 + \ln {x_2}}}{{({x_2} - {x_1})}} + \frac{{{x_1}\ln {x_1}}-{{x_2}\ln {x_2}}}{{{{({x_2} - {x_1})}^2}}} \Big],\\
&&{I_2}(\textit{x}_1 , x_2 ) = \frac{1}{{16{\pi ^2}}}\Big[ - \frac{{1 + \ln {x_1}}}{{({x_2} - {x_1})}} - \frac{{{x_1}\ln {x_1}}-{{x_2}\ln {x_2}}}{{{{({x_2} - {x_1})}^2}}} \Big],\\
&&{I_3}(\textit{x}_1 , x_2 ) = \frac{1}{{32{\pi ^2}}}\Big[  \frac{{3 + 2\ln {x_2}}}{{({x_2} - {x_1})}} - \frac{{2{x_2} + 4{x_2}\ln {x_2}}}{{{{({x_2} - {x_1})}^2}}} -\frac{{2x_1^2\ln {x_1}}}{{{{({x_2} - {x_1})}^3}}}  +  \frac{{2x_2^2\ln {x_2}}}{{{{({x_2} - {x_1})}^3}}}\Big], \\
&&{I_4}(\textit{x}_1 , x_2 ) = \frac{1}{{96{\pi ^2}}} \Big[ \frac{{11 + 6\ln {x_2}}}{{({x_2} - {x_1})}}- \frac{{15{x_2} + 18{x_2}\ln {x_2}}}{{{{({x_2} - {x_1})}^2}}} + \frac{{6x_2^2 + 18x_2^2\ln {x_2}}}{{{{({x_2} - {x_1})}^3}}}  \nonumber\\
&&\hspace{2.1cm} + \: \frac{{6x_1^3\ln {x_1}}-{6x_2^3\ln {x_2}}}{{{{({x_2} - {x_1})}^4}}}  \Big], \\
&& {G_1}(\textit{x}_1 , x_2 , x_3) \nonumber\\
&&\hspace{0.6cm} =  \frac{1}{{16{\pi ^2}}}\Big[ \frac{{{x_1}\ln {x_1}}}{{({x_1} - {x_2})({x_1} - {x_3})}} + \frac{{{x_2}\ln {x_2}}}{{({x_2} - {x_1})({x_2} - {x_3})}}  +  \frac{{{x_3}\ln {x_3}}}{{({x_3} - {x_1})({x_3} - {x_2})}}\Big], \\
&& {G_2}(\textit{x}_1 , x_2 , x_3) \nonumber\\
&&\hspace{0.6cm} =  \frac{1}{{16{\pi ^2}}}\Big[  \frac{{x_1^2\ln {x_1}}}{{({x_1} - {x_2})({x_1} - {x_3})}} +\frac{{x_2^2\ln {x_2}}}{{({x_2} - {x_1})({x_2} - {x_3})}} +   \frac{{x_3^2\ln {x_3}}}{{({x_3} - {x_1})({x_3} - {x_2})}} \Big], \quad \\
&& {G_3}(\textit{x}_1 , x_2 , x_3, x_4) \nonumber\\
&&\hspace{0.6cm} = \frac{1}{{16{\pi ^2}}}\Big[\frac{{{x_1}\ln {x_1}}}{{({x_1} - {x_2})({x_1} - {x_3})({x_1} - {x_4})}}  + \frac{{{x_2}\ln {x_2}}}{{({x_2} - {x_1})({x_2} - {x_3})({x_2} - {x_4})}} \nonumber\\
&&\hspace{0.6cm}\quad + \frac{{{x_3}\ln {x_3}}}{{({x_3}  - {x_1})({x_3} - {x_2})({x_3} - {x_4})}} + \: \frac{{{x_4}\ln {x_4}}}{{({x_4} - {x_1})({x_4} - {x_2})({x_4} - {x_3})}}\Big] , \\
&&{G_4}(\textit{x}_1 , x_2 , x_3, x_4) \nonumber\\
&&\hspace{0.6cm} = \frac{1}{{16{\pi ^2}}}\Big[\frac{{x_1^2\ln {x_1}}}{{({x_1} - {x_2})({x_1} - {x_3})({x_1} - {x_4})}}   + \frac{{x_2^2\ln {x_2}}}{{({x_2} - {x_1})({x_2} - {x_3})({x_2} - {x_4})}} \nonumber\\
&&\hspace{0.6cm}\quad + \frac{{x_3^2\ln {x_3}}}{{({x_3}  - {x_1})({x_3} - {x_2})({x_3} - {x_4})}}   +  \frac{{x_4^2\ln {x_4}}}{{({x_4} - {x_1})({x_4} - {x_2})({x_4} - {x_3})}}\Big].
\end{eqnarray}

\end{document}